\documentclass[%
superscriptaddress,
 amsmath,amssymb,
aip,
]{revtex4-2}

\usepackage[utf8]{inputenc}
\usepackage[T1]{fontenc}
\usepackage{chemformula}
\usepackage{siunitx}
\usepackage{appendix}
\usepackage{graphicx}
\usepackage{dcolumn}
\usepackage{bm}
\usepackage{float}
\usepackage{array, makecell}
\usepackage[mathlines]{lineno}
\usepackage{comment}

\UseRawInputEncoding

\begin{document}


\title[]{Oscillating reaction in porous media under saddle flow}
\author{Satoshi Izumoto}
\email{satoshi.izumoto@hotmail.co.jp}
\affiliation{ 
Institut de Physique de Rennes, Univ. Rennes 1, CNRS, Unit\'e Mixte de Recherche 6118, Rennes, France
}%

%

\begin{abstract}
{Pattern formation due to oscillating reactions represents variable natural and engineering systems, but previous studies employed only simple flow conditions such as uniform flow and Poiseuille flow. We studied the oscillating reaction in porous media, where dispersion enhanced the spreading of diffusing components by merging and splitting flow channels. We considered the saddle flow, where the stretching rate is constant everywhere. We generated patterns with the Brusselator system and classified them by instability conditions and P\'eclet number (Pe), which was defined by the stretching rate. The results showed that each pattern formation was controlled by the stagnation point and stable and unstable manifolds of the flow field due to the heterogeneous flow fields and the resulting heterogeneous dispersion fields. The characteristics of the patterns, such as the position of stationary waves parallel to the unstable manifold and the size of local stationary patterns around the stagnation point, were also controlled by Pe.}
\end{abstract}


\maketitle


\section{Introduction}
Synchronization of the individual oscillating components and subsequent wave and pattern formation are found in many systems in physics \citep{Cross1993,Haudin2009,VanSaarloos2003}, biology \citep{Bressloff2013,Beta2017}, chemistry \citep{Sagues2003,Rotermund2009} and earth science \citep{Samuelson2019,Martin2003a}. The oscillating chemical reaction, such as the Belousov-Zhabotinsky reaction, is a prominent example and has been widely studied \citep{nicolis1977}. Previous studies investigated how pattern formations are influenced in various conditions, such as under advective flow \citep{Doan2018,Nevins2016,Nugent2004,Vasquez2004,Neufeld2001,Escala2014,Schwartz2008}, with anomalous diffusion and cross diffusion \citep{Manz2004,Golovin2008,Berenstein2013,Paoletti2006} and with spatial heterogeneity of kinetic parameters and boundary conditions \citep{Ginn2004a,Kozak2019,Wang2017,klika2018}. These fundamental studies led to the engineering application of oscillating reactions, including chemical computation processors \citep{Ryzhkov2021,Parrilla-Gutierrez2020,Wang2016}, mechanical oscillation \citep{Yoshida1996,Homma2017,Li2019,Mallphanov2020}, viscosity oscillation \citep{Ahmadi2020}, and separation of colloids \citep{Cui2020}. In addition to these engineering applications, the oscillating reactions are experimentally applied to mimic biological systems to unravel the underlying mechanisms \citep{Litschel2018,Ren2017,Ren2020}. 
However, most of the studies employed simple geometry and simple flow conditions, such as uniform flow and Poiseuille flow. Studying oscillating reactions in other geometries is important for extending the application of oscillating reactions. For example, recent studies investigated the pattern formation on curved domains that deform dynamically \citep{Nishide2022,Frank2019,Krause2021}. Such system is closely related to pattern formations in biological systems such as cell membrane and growing organ \citep{Nishide2022}. \\
In this context, we focus on oscillating reaction in porous media. Porous media is ubiquitous in many natural and engineering systems. It represents the environment for bacterial habitat in subsurface \citep{Coyte2017,DeAnna2021,Dehkharghani2019} and heterogeneous flow fields due to obstacles placed in the flow field \citep{Atis2013}. Thus, studying oscillating reactions in porous media would give insight into the predator-prey systems, that may form the wave front similar to that in oscillating reactions \citep{Goldbeter2017,Goldbeter2018}, under subsurface and in open flows including obstacles. Furthermore, the mixing of chemical species are influenced by the complex flow path in porous media, that may result in new patterns. This would broaden the engineering application of oscillating reaction.\\
Many studies investigated reactive transport in porous media, including bimolecular reactions \citep{DeAnna2014a,Valocchi2019b,Yoon2021}, precipitation/dissolution reactions \citep{Izumoto2022,Soulaine2018} in the context of geoscience and chemical engineering. However, only a few previous studies investigated the oscillating reaction under conditions relevant to porous media. \citet{Sepulchre1993} simulated the translational motion of spiral waves in the presence of an impermeable obstacle. \citet{Marlow1997} numerically investigated how convection influences the Turing pattern by introducing Darcy's law. \citet{Ginn2005} experimentally observed the spiral waves under nonexcitable obstacles. More recently, \citet{Krause2021} numerically and theoretically investigated stationary pattern formations in complex geometries of boundaries. Even though the conditions in these studies are partly relevant to those in porous media, none of them focused on the oscillation reaction in porous media under flow. There are a series of previous studies that investigated reactive transport of autocatalytic fronts in porous media, which reported the static wave front against the advection and microscopic effects when the pores are of the size of the chemical front width \citep{Atis2012,Atis2013,Chevalier2017}. Such findings are potentially applicable to the oscillating reaction because the oscillating reaction is the combination of the autocatalytic species and the inhibitor species.\\
One of the characteristics of porous media is the spreading of solutes in space due to the branching and converging of flow channels. This effect, called dispersion, is included in the advection-dispersion-reaction equation:
\begin{equation}
\label{eq1}
    \frac{\partial C}{\partial t} = -v\cdot\nabla C+\nabla\cdot(D(v)\nabla C)+R
\end{equation}
where $C$ is the volume-averaged concentration of a chemical species, $v$ is the local velocity, $D(v)$ is the dispersion tensor as a function of local velocity, and $R$ is the reaction term. The above equation does not consider the variation of concentration within the pore space, but rather uses the averaged concentration in a representative elementary volume of the porous medium, which includes a large number of grains and pores \citep{Rolle2019,Dentz2011}. This approximation is valid when the concentration varies more steeply in a larger volume than in a pore-scale.
In a uniform flow in porous media, it would be rather straight-forward to apply the result obtained using the advection-diffusion-reaction equation because the dispersion is constant everywhere. However, the reactive transport in porous media is very different in a heterogeneous flow field because the dispersion becomes heterogeneous. One of the prototypes of such a flow field is saddle flow, where the stretching rate of the fluid flow is constant everywhere. This flow field is reminiscent of flows in nature, including a stagnation point and stable/unstable manifolds of the flow \citep{Haller2015LagrangianStructures}. These manifolds work as transport barriers for passive and active tracers \citep{Mahoney2012,Mitchell2012,Berman2021TransportFlows}. Also, the chemical species are strongly mixed due to the strong deformation of fluid elements around the stagnation point, which leads to an enhanced reaction \citep{izumoto2023b}.\\
Here, we studied oscillating reactions in porous media under saddle flow. We employed the Brusselator, which is one of the typical nonlinear reaction kinetics used to study oscillating phenomenon \citep{nicolis1977}. First, we theoretically show the predictions of possible patterns. Next, we show the results of simulations only with diffusion (with-diffusion cases hereafter) and with diffusion and dispersion (with-dispersion cases hereafter) cases. The with-diffusion cases give insights into the oscillating reaction at the pore scale with stable and unstable manifolds in the flow fields, which represent backbones of the flow field in pore space \citep{Metcalfe2022}. The with-dispersion cases represent continuum-scale reactive transport in porous media, where the concentrations are defined by volume averaging.

\section{Theory}
First, we analytically investigated the patterns in the saddle flow with dispersion. We employed the irreversible Brusselator as the reaction term:
\begin{equation}
\label{eq2}
    \begin{gathered}
    A \xrightarrow{\mathit{k1}} X \\
    B+X \xrightarrow{\mathit{k2}} Y+D \\
    2X+Y \xrightarrow{\mathit{k3}} 3X \\
    X \xrightarrow{\mathit{k4}} E \\
    \end{gathered}
\end{equation}
where $k_i$, $i\in[1,4]$ are the rate constants, A and B are the initial reactants, X is the autocatalytic species and Y is the inhibitor species. We assumed homogeneous A and B and small $k_1$ and $k_2$ (pool approximation). The corresponding reaction equations can be written as:
\begin{equation}
\label{eq3_0}
    \frac{\partial \tilde{X}_{aq}}{\partial \tilde{t}} = k_1\tilde{A}_{aq}-\left(k_2\tilde{B}_{aq}+k_4\right)\tilde{X}_{aq}+k_3\tilde{X}_{aq}^2\tilde{Y}_{aq}
\end{equation}

\begin{equation}
\label{eq4_0}
    \frac{\partial \tilde{Y}_{aq}}{\partial \tilde{t}} = -k_2\tilde{B}_{aq}\tilde{X}_{aq}-k_3\tilde{X}_{aq}^2\tilde{Y}_{aq}
\end{equation}
where, for example $\tilde{A}_{aq}$, represents the concentration of the species $A$ in the aqueous phase. This is the case for the pure aqueous solution. In porous media, we consider the volume-averaged concentrations in a representative elementary volume that includes many numbers of grains and pores. Thus, the concentrations of chemical species in the aqueous phase, for example $\tilde{A}_{aq}$, should be replaced by the volume-averaged quantity $\omega\tilde{A}_{aq}$, where $0<\omega<1$ is the porosity of the porous medium defined by the proportion of the aqueous phase. We replace \{$\tilde{A}_{aq},\tilde{B}_{aq},\tilde{X}_{aq},\tilde{Y}_{aq}$\} by \{$\omega\tilde{A}_{aq},\omega\tilde{B}_{aq},\omega\tilde{X}_{aq},\omega\tilde{Y}_{aq}$\} in Eq.\ref{eq3_0},\ref{eq4_0}, and divide by $\omega$, which leads to:
\begin{equation}
\label{eq3_1}
    \frac{\partial \tilde{X}_{aq}}{\partial \tilde{t}} = k_1\tilde{A}_{aq}-\left(\omega k_2\tilde{B}_{aq}+k_4\right)\tilde{X}_{aq}+\omega^2 k_3\tilde{X}_{aq}^2\tilde{Y}_{aq}
\end{equation}

\begin{equation}
\label{eq4_1}
    \frac{\partial \tilde{Y}_{aq}}{\partial \tilde{t}} = -\omega k_2\tilde{B}_{aq}\tilde{X}_{aq}-\omega^2 k_3\tilde{X}_{aq}^2\tilde{Y}_{aq}
\end{equation}
By comparing Eq.\ref{eq3_0},\ref{eq4_0} and Eq.\ref{eq3_1},\ref{eq4_1}, we found that we can recover the original Brusselator system (Eq.\ref{eq2},\ref{eq3_0},\ref{eq4_0}) from the Brusselator system in porous media (Eq.\ref{eq3_1},\ref{eq4_1}) if we replace \{$\omega k_2$,$\omega^2 k_3$\} by \{$k_2$,$k_3$\}. Thus, the Brusselator system in porous media can be defined by modifying the reaction constants as a function of porosity. We use the original Brusselator sytem (Eq.\ref{eq2}) hereafter to avoid confusion without loss of generality.\\
The reactive transport of the solutes was governed by the 2D advection-dispersion-reaction equation:
\begin{equation}
\label{eq3}
    \frac{\partial \tilde{X}}{\partial \tilde{t}} = -\tilde{v}\cdot\nabla \tilde{X}+\nabla \cdot\left(\tilde{D}_{disp,X}\nabla \tilde{X}\right)+k_1\tilde{A}-\left(k_2\tilde{B}+k_4\right)\tilde{X}+k_3\tilde{X}^2\tilde{Y}
\end{equation}

\begin{equation}
\label{eq4}
    \frac{\partial \tilde{Y}}{\partial \tilde{t}} = -\tilde{v}\cdot\nabla \tilde{Y}+\nabla \cdot\left(\tilde{D}_{disp,Y}\nabla \tilde{Y}\right)+k_2\tilde{B}\tilde{X}-k_3\tilde{X}^2\tilde{Y}
\end{equation}
where $\tilde{v}$ is fluid velocity in continuum-scale (defined by $\tilde{v}=\omega \tilde{q}$ where $\tilde{q}$ is the pore-scale velocity), $\tilde{D}_{disp,i}$ is the dispersion tensor and the concentrations are volume-averaged. The dispersion tensor $\tilde{D}_{disp,i}$, where $i \in \{X,Y\}$, can be approximated as the sum of molecular diffusion and the mechanical dispersion due to the flow channeling in porous media as:
\begin{equation}
\label{eq5}
 \tilde{D}_{disp,i} = (\tilde{D}_{m,i} + \tilde{\alpha} |\tilde{{v}|}) I
\end{equation}
$\tilde{D}_{m,i}$ is the molecular diffusion coefficient, $\tilde{\alpha}$ is dispersivity, which is an intrinsic property of the porous medium (normally assumed to be proportional to grain size), and $I$ is the identity matrix \citep{Yortsos1988,Ghesmat2008}. If the solutes are initially distributed in a narrow region within pore space, the velocity dependency of the dispersion tensor changes over time until it reaches the asymptotic regime. This transition from a preasymptotic to an asymptotic regime is achieved for a sufficiently long time or large length by allowing full sampling of the flow field heterogeneity \citep{Dentz2011}. In Eq.\ref{eq5}, we assumed an asymptotic regime. This is valid when the reaction rates are small enough to keep the spread of chemical species. Note that we also studied the cases when the oscillating reaction occurs within pore space and the assumptions for continuum-scale porous media do not hold. This was achieved by setting $\alpha=0$. Such cases, called with-diffusion cases, clarify how pattern formation is controlled by the stable and unstable manifolds within the pore space \citep{Metcalfe2022}. \\
The flow field was saddle flow that included the stagnation point at $(x,y)=(0,0)$ (Fig.\ref{figCalc}a):
\begin{equation}
\label{eq6}
 (\tilde{v}_{x},\tilde{v}_{y})=(\tilde{\gamma}\tilde{x},-\tilde{\gamma}\tilde{x})
\end{equation}
We introduced nondimentional variables: \{$t=\tilde{t}/t_c,\enskip x = \tilde{x}/l_c, y = \tilde{y}/l_c,\enskip X=\tilde{X}/\Bar{X},\enskip Y=\tilde{Y}/\Bar{Y},\enskip D_{m,X} = \tilde{D}_{m,X}/D_c,\enskip D_{m,Y} = \tilde{D}_{m,Y}/D_c, \enskip \alpha=\tilde{\alpha}/l_c, \enskip \gamma=\tilde{\gamma}t_c,\enskip v=\tilde{v}/v_c$\}. The characteristic scales were determined based on the reaction rates, diffusion coefficient, and length scale as \{$l_c=(D_ct_c)^{1/2},\enskip t_c=1/k_4,\enskip v_c=l_c/t_c,\enskip \Bar{X}=\Bar{Y}=(k_4/k_3)^{1/2}$\}. We set $D_c$ to 1. The non-dimentionalized forms of Eq.\ref{eq3},\ref{eq4} are:
\begin{equation}
\label{eq7}
    \frac{\partial X}{\partial t} = -v\cdot\nabla X+\nabla \cdot\left(D_{disp,X}\nabla X\right)+A-(B+1)X+X^2Y
\end{equation}
\begin{equation}
\label{eq8}
    \frac{\partial Y}{\partial t} = -v\cdot\nabla Y+\nabla \cdot\left(D_{disp,Y}\nabla Y\right)+BX-X^2Y
\end{equation}
with the velocity field:
\begin{equation}
\label{eq9}
 (v_x,v_y)=(\gamma x,-\gamma y)
\end{equation}
and the dispersion tensor:
\begin{equation}
\label{eq10}
 D_{disp,i} = (D_{m,i} + \alpha |v|) I
 = (D_{m,i} + \alpha \gamma \sqrt{x^2+y^2}) I
\end{equation}

When the velocity $v$ and its gradient $\nabla v$ are small enough, the above equations can be simplified by dropping the advection term and replacing the dispersion tensor by the dispersion coefficient $\hat{D}_{disp,i}=D_{m,i}+\alpha |v|$ as:
\begin{equation}
\label{eq11}
    \frac{\partial X}{\partial t} = \hat{D}_{disp,X}\nabla^2X+A-(B+1)X+X^2Y
\end{equation}
\begin{equation}
\label{eq12}
    \frac{\partial Y}{\partial t} = \hat{D}_{disp,Y}\nabla^2Y+BX-X^2Y
\end{equation}
If the concentrations are homogeneous in the entire field, Eq.\ref{eq11} and Eq.\ref{eq12} have the steady state solution (X,Y) = ($A$,$B/A$). This solution is unstable toward the Hopf instability if $B>B_H = 1+A^2$, which induces homogeneous oscillation of X and Y in the entire field. If the concentrations are heterogeneous but in the absence of porous media, we have $\hat{D}_{disp,i}=D_{m,i}$. This case has been extensively studied \citep{nicolis1977}, and it is known that the Hopf instability may lead to traveling waves. When $B>B_T = (1+A\sqrt{\delta})^2$, where $\delta=\hat{D}_{m,X}/\hat{D}_{m,Y}$, the solution is unstable toward Turing instability. This leads to a stationary pattern called the Turing pattern. When $B<B_H$ and $B<B_T$, X and Y are homogeneous and stationary. When $B>B_H$ and $B>B_T$, the pattern is governed by the Hopf instability if $B_H<B_T$, and by the Turing instability if $B_T<B_H$. If the concentrations are heterogeneous in the presence of porous media, it is necessary to calculate $\hat{D}_{disp,i}$ to classify the instability conditions. Considering the formula for dispersion coefficient (Eq.\ref{eq10}) and the velocity field (Eq.\ref{eq9}), we obtain $B_T$:
\begin{equation}
\label{eq13}
    B_T = \left(1+A\sqrt{\frac{D_{m,X} + \alpha \gamma \sqrt{x^2+y^2}}{D_{m,Y} + \alpha \gamma \sqrt{x^2+y^2}}}\right)^2
\end{equation}
We calculated $B_T$ on the x-axis ($y=0$). We fixed $(A,D_{m,X},D_{m,Y})=(4.5,1,8)$ so that both Turing and Hopf instability conditions were included in the computational domain by varying $\gamma\alpha\in\{0.02,0.04,0.08,0.16,0.32\}$. $B_H$ was also calculated with the same value ($A=4.5$). The results suggested four patterns (Fig.\ref{figCalc}b): (I) Homogeneous steady state when $B<B_T$ and $B<B_H$; (II) Local Turing instability around the stagnation point when $B_T<B$ and $B<B_H$; (III) Turing instability around the stagnation point and Hopf instability away from the stagnation point when $B>B_T$, $B>B_H$ and $B_T$ crosses $B_H$ and (IV) Hopf instability when $B>B_H$ and $B_H<B_T$. The position of the edge of the Turing instability zone should be given by $B=B_{T}$ in the case of (II) and by $B_{T}=B_{H}$ in the case of (III). By solving Eq.\ref{eq13} under these conditions, the edge position of the Turing instability zone can be written as:
\begin{equation}
\label{eq14}
    \sqrt{x^2+y^2}=\frac{D_{m,Y}\delta_c-D_{m,X}}{\alpha\gamma(1-\delta_c)}
\end{equation}
where $\delta_c$ corresponds to the critical ratio of the diffusion coefficient in a homogeneous system calculated by $\delta_c = ((\sqrt{B}-1)/A_0)^2$ in the case of (II) and $\delta_c = ((\sqrt{1+A^2}-1)/A)^2$ in the case of (III) \citep{Budroni2016a}. The above equation shows that the edge should have a circular shape. We further considered the parameter called supercriticality, $\varepsilon = (B-B_c)/B_c$ where $B_c=\min\{B_H,B_T\}$  \citep{Pena2001}, which quantifies how far the system is from the equilibrium state. It has been reported that the type of Turing pattern (hexagon, stripes, and reentrant hexagon) changes with $\varepsilon$ \citep{Pena2001}. Thus, different types of Turing patterns should appear in porous media because $\varepsilon$ varies over space. 
\\
When the advection $v$ is large, Eq.\ref{eq11},\ref{eq12} do not hold. Previous studies suggested the theoretical framework to analyse the pattern formation under the advection field by considering the effect of boundary conditions \citep{Andresen1999,Bamforth2000,Kaern2000,McGraw2005,klika2018}, highlighting that homogeneous perturbation can not grow close to the boundary \citep{klika2018}. However, we could not apply these methods to our system because both the flow field and the dispersion field are heterogeneous. Thus, we investigated these cases mainly through simulations.\\

\section{Simulation methods}
\begin{figure}
    \centering
    \includegraphics{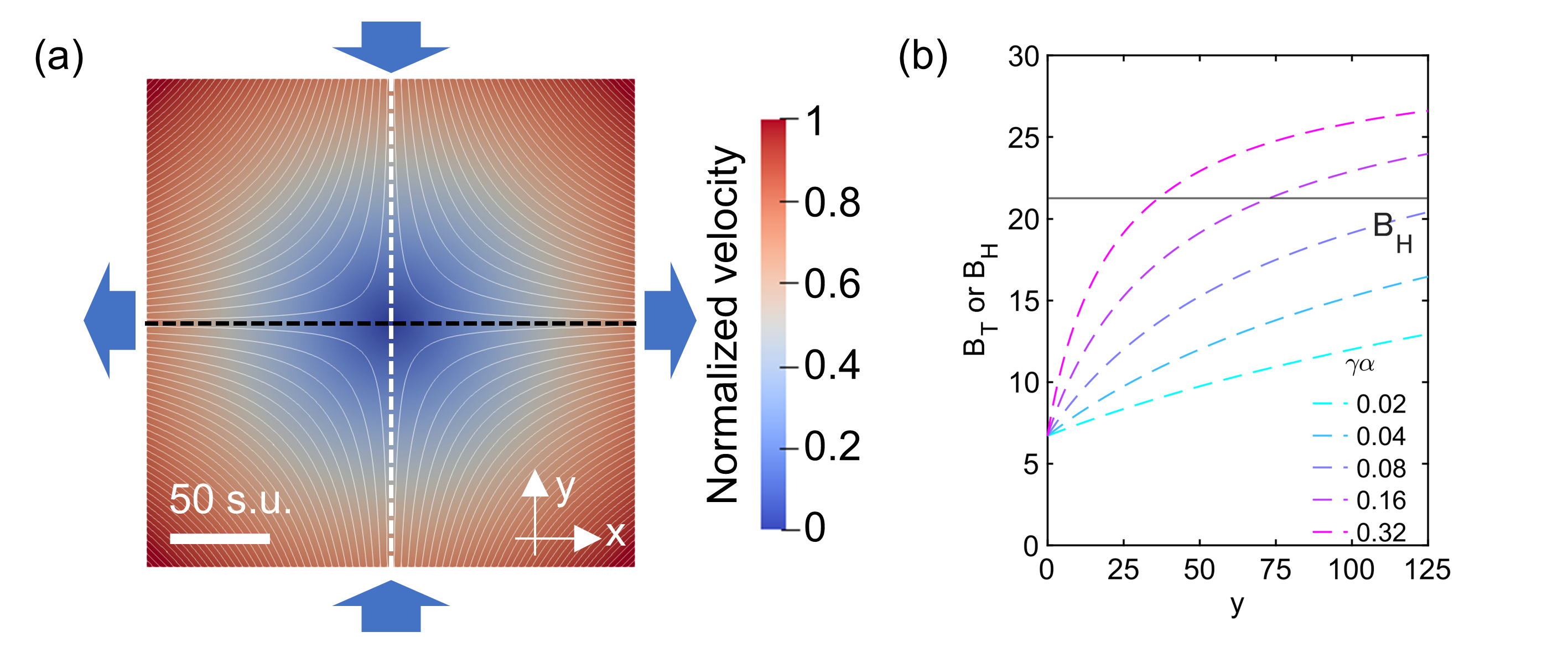}
    \caption{(a) The streamlines and velocity magnitude in saddle flow normalized by the maximum velocity. The inflow and outflow directions are indicated by the blue arrows. The entire zone is 250 by 250 space units (s.u.), with $x\in [-125,125]$ and $y\in [-125,125]$. The stagnation point is at the center. The x=0 line corresponds to the stable manifold (white dotted line), and the y=0 line corresponds to the unstable manifold (black dotted line). (b) The analytical solutions of $B_{T}$ (the dotted lines) and $B_{H}$ (the solid line) as a function of space along $x=0$ and $y\in\{0,125\}$, where $(x,y)=(0,0)$ is the stagnation point. The parameters were $(A,D_{m,X},D_{m,Y})=(4.5,1,8)$ and $\gamma\alpha\in\{0.02,0.04,0.08,0.16,0.32\}$.}.
    \label{figCalc}
\end{figure}

We have run numerical simulations using the open-source CFD software OpenFOAM, which utilizes the finite volume method. The governing equations were the same as in the theory (Eq.\ref{eq7}-\ref{eq10}). We considered the square domain of the size $L=250$, where $x\in[-125,125]$, $y\in[-125,125]$ (Fig.\ref{figCalc}a). Initially, A and B homogeneously spread in the domain $(A,B,X,Y)=(A_0,B_0,0,0)\enskip \forall x,y$. As for the inlet boundary condition at $y=\pm L/2$, we imposed the zero gradient condition for X and Y, and constant concentrations for A and B as $A_0$ and $B_0$. For the outlet at $x=\pm L/2$, we imposed the zero gradient condition for all the species. The temporal resolution was 0.001 time unit and the spatial resolution was 0.625 space unit, which were similar to the previous study employing Brusselator for the diffusion-reaction problem \citep{Budroni2016a}. We used the Euler method as a temporal discretisation scheme and the linear interpolation scheme for face-centered values from cell-centered values. We first simulated diffusion cases by setting the dispersivity to zero for with-diffusion cases ($\alpha=0$). This represented the typical pore-scale flow field in 3D porous media, where stable and unstable manifolds induce exponential stretching of fluid elements \citep{Metcalfe2022}. There was either Turing instability or Hopf instability in the entire domain because the diffusion was constant everywhere. We established Turing and Hopf instabilities by appropriately setting $A_0, B_0, D_{m,X}, D_{m,Y}$ (Table \ref{tableParams1}). We varied the stretching rate $\gamma$ in each case to investigate how the advection changes the pattern formation. The corresponding P\'eclet numbers (Pe) were calculated by $Pe=L/2\times\gamma/((D_{m,X}+D_{m,Y})/2)=L\gamma/(D_{m,X}+D_{m,Y})$. Next, we simulated with-dispersion cases ($\alpha>0$), which represented the continuum-scale reactive transport in porous media. Following the theoretical prediction, we distinguished the cases by types of instabilities: Turing instability, mixture of Turing and Hopf instability, and Hopf instability. We changed $A_0, B_0, D_{m,X}, D_{m,Y}, \alpha$ and $\gamma$ to establish these patterns (Table \ref{tableParams2}). For each pattern, we varied $\gamma$ while keeping $\alpha\gamma$ constant to investigate the effect of advection on pattern formation (Eq.\ref{eq9},\ref{eq10}). Note that if we only change $\gamma$, not only the advection term but also the dispersion term $D_{disp,i}$ changes following Eq.\ref{eq10}. In this case, the effects of advection and dispersion are coupled. To avoid this, we also modified $\alpha$ so that $\gamma\alpha$ did not change. This ensures that the terms that included $D_{disp,i}$ did not change so that the effect was only due to the terms $v\cdot\nabla X$ and $v\cdot\nabla Y$ in Eq.\ref{eq7},\ref{eq8}.\\

\begin{table}
\centering
\caption{\label{tableParams1}Fixed parameters $(A_0,B_0,D_{m,X},D_{m,Y})$ and the variable parameter $\gamma$ in the with-diffusion cases ($\alpha=0$).}

\begin{tabular}{c c c}
\hline
&Turing&Hopf\\ \hline
 $(A_0,B_0,D_{m,X},D_{m,Y})$&$(4.5,13.3,1,8)$&$(1,2.2,1,0.1)$ \\
$\gamma \times 10^{3}$ range&[$0.592,94.7$] &[$59.2,592$]\\
Pe range& [0.0164, 2.63] & [13.5,135] \\
\hline
\end{tabular}
\end{table}

\begin{table*}
\caption{\label{tableParams2}Fixed parameters $(A_0,B_0,D_{m,X},D_{m,Y},\gamma\alpha)$ and the variable parameter $\gamma$ in the with-dispersion cases ($\alpha > 0$).}

\begin{tabular}{c c c c}
\hline
&Turing&Turing+Hopf&Hopf\\ \hline
 $(A_0,B_0,D_{m,X},D_{m,Y},\gamma\alpha)$ &$(4.5,13.3,1,8,0.0592)$&$(4.5,33,1,8,0.237)$& $(1,2.2,1,0.1,0.444)$ \\
$\gamma \times 10^{3}$ range&[$5.92,296$]&[$5.92,503$]&[$296,888$]\\
Pe range & [0.164, 8.2]& [0.164, 14] & [68, 202]\\
\hline
\end{tabular}
\end{table*}

\section{Results and discussions}

\subsection{Low Pe regime}
\begin{figure}
    \centering
    \includegraphics[width=\textwidth,height=\textheight,keepaspectratio]{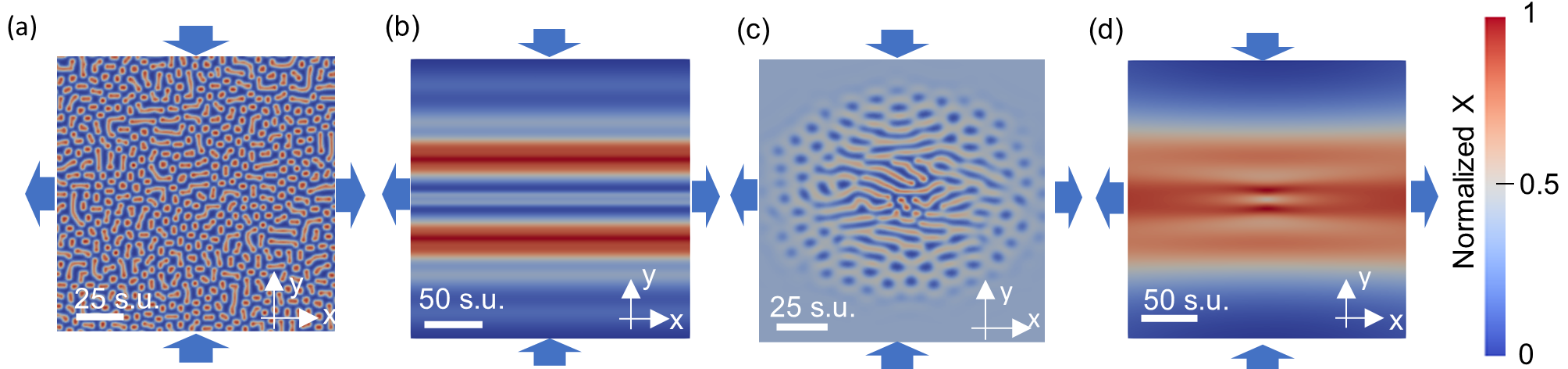}
    \caption{A snapshot of normalized concentration field of the species X in low Pe. The normalized concentration is defined by $(X-min(X)/(max(X)-min(X))$. All the figures share the same color bar. The blue arrows indicate the flow direction. (a) Turing instabilities in the with-diffusion case with Pe = 0.0164 close to the stagnation point. (Multimedia view) (b) Hopf instabilities in the with-diffusion case with Pe = 26.9. (Multimedia view) (c) Turing instabilities in the with-dispersion case with Pe = 0.82. (Multimedia view) (d) Hopf instability conditions in the with-dispersion case with Pe = 68. (Multimedia view) }
    \label{figLowPe}
\end{figure}

\begin{figure}
    \centering
    \includegraphics[width=\textwidth,height=\textheight,keepaspectratio]{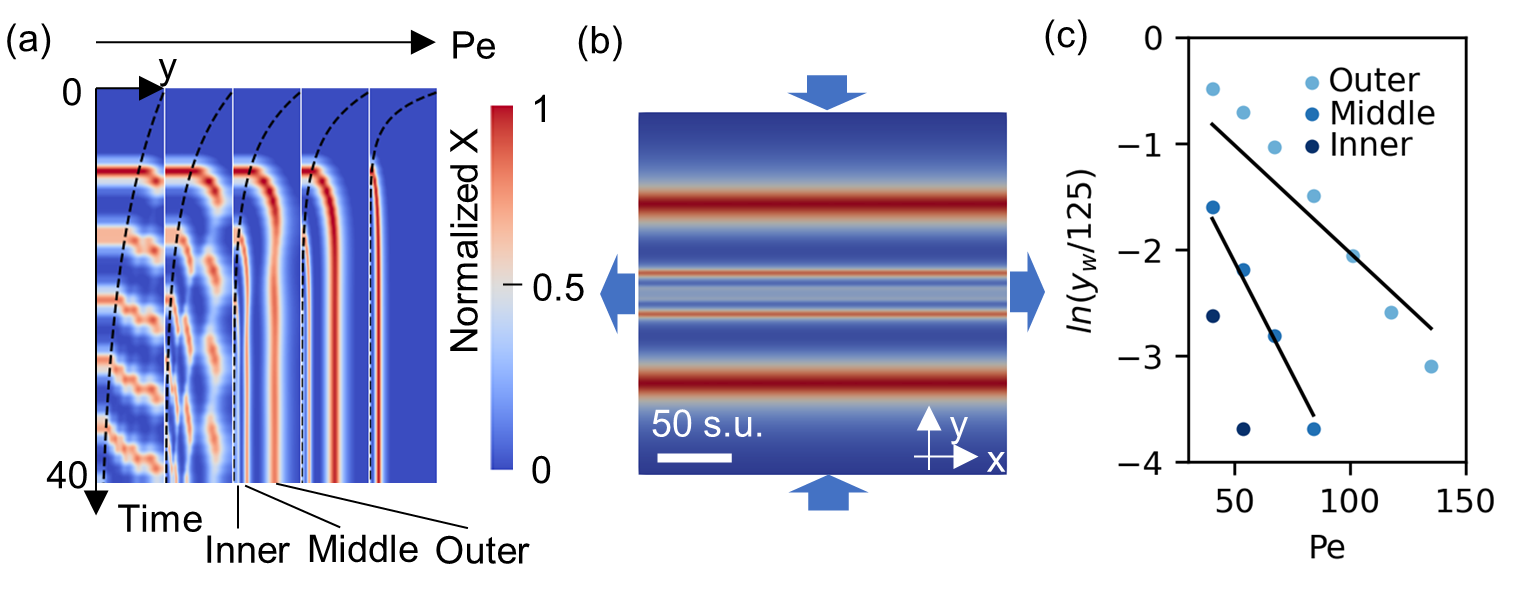}
    \caption{Hopf instabilities in with-diffusion cases in high Pe. (a) S-T map of the normalized concenetration of species X in different $Pe\in\{13.5,26.9,40.4,53.8,101\}$, at $x = 0$, $y\in[0,125]$. The dotted lines represent the advection of the fluid element that starts at the boundary at time 0. The normalized concentration is defined by $(X-min(X)/(max(X)-min(X))$.  (b) A snapshot of the normalized concentration field X, showing a stationary pattern in high Pe (= 40.4). The blue arrows indicate the flow direction. The space unit is abbreviated to s.u.. (a) and (b) share the same range of concentration. (Multimedia view) (c) $ln(y_m/125)$ of the stationary wave as a function of $Pe$, where $y_m$ (the position of the wave in y coordinate) is divided by the size of the domain (125). The "Outer", "Middle" and "Inner" corresponds to each stationary waves indicated in (a). The black lines are the fitted functions $ln(y_m/125)=a\: Pe$, where $a$ is the fitting parameter. For the "Outer" case, $a=-0.0203$ with $R^2=0.976$. For the "Middle" case, $a=-0.0423$ with $R^2=0.999$.}
    \label{figNoDispHopf}
\end{figure}

In this section we discuss the results obtained in the low Pe regime. We first show the results of with-diffusion cases ($\alpha=0$). For the Turing instability condition at low Pe, we observed Turing patterns slightly moving following the advection (Fig.\ref{figLowPe}a, (Multimedia view)). For the Hopf instability condition at low Pe, we observed plane traveling waves staring from the unstable manifold (Fig.\ref{figLowPe}b, (Multimedia view)). The space-time map (S-T map) of the lowest Pe shows that the homogeneous oscillation was gradually delayed from the inlet boundary toward the unstable manifold (Fig.\ref{figNoDispHopf}a, left panel). The timing of the disturbance matched the arrival of the advected fluid element starting at the boundary at t = 0 (dotted line in Fig.\ref{figNoDispHopf}a, left panel). The oscillation at the unstable manifold was least affected by the boundary because the fluid element did not reach the unstable manifold, where $v_y = 0$. Thus, the unstable manifold became the start of the traveling waves. A similar mechanism was reported by \citet{klika2018}, where the advection suppresses the growth of the perturbation close to the inlet boundary.

\begin{figure}
    \centering
    \includegraphics[width=\textwidth,height=\textheight,keepaspectratio]{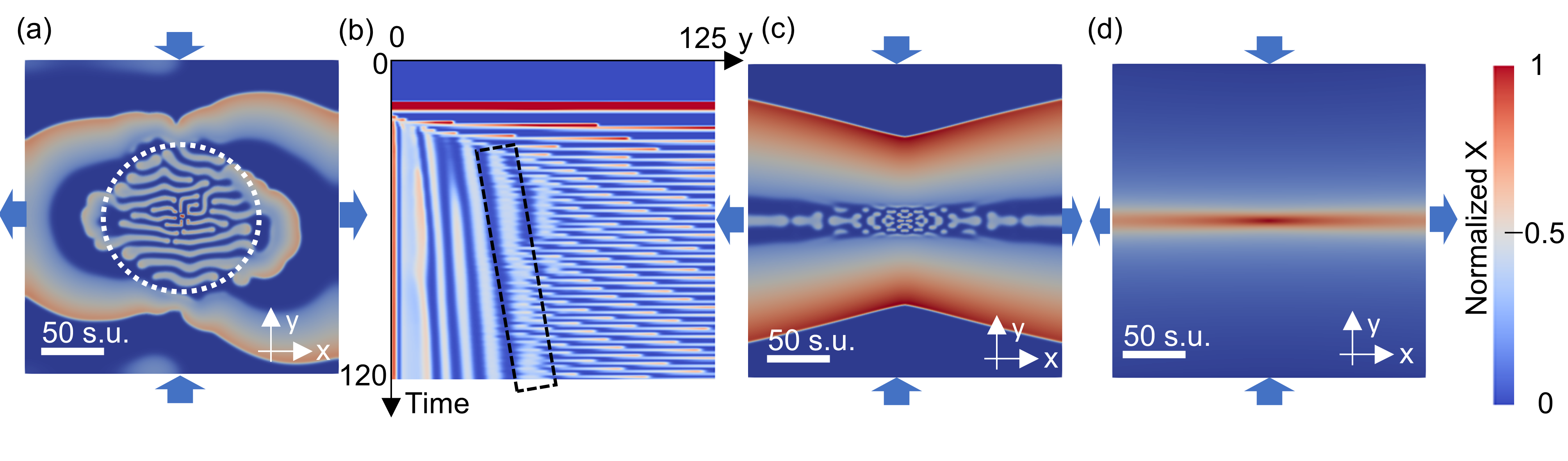}
    \caption{Turing and Hopf instabilities in with-dispersion cases. We show the normalized concentration of the species X defined by $(X-min(X)/(max(X)-min(X))$. The blue arrows indicate the flow direction. (a) Turing instability close to the stagnation point and Hopf instability far from the stagnation point in low Pe (= 0.164). The white dotted line indicates the interface of the Turing pattern and the traveling wave zone determined by the simulation. (Multimedia view) (b) S-T map at $x = 0$, $y\in[0,125]$. The black dotted rectangle indicates the track of a part of the Turing pattern advected by the flow. (c) Turing instability close to the stagnation point and Hopf instability far from the stagnation point in middle Pe (= 1.32). (Multimedia view) (d) Stationary pattern in high Pe (= 1.4). The abbreviation s.u. represents a space unit. }
    \label{figPorousTuringHopf}
\end{figure}

Next, we discuss the patterns in with-dispersion cases ($\alpha>0$). In low Pe, we observed three types of patterns as expected from the theory: localized Turing patterns, coexistence of Turing patterns and traveling waves, and only traveling waves. In the localized Turing pattern (Fig.\ref{figLowPe}c, (Multimedia view)), we observed hexagons, stripes, and reentrant hexagons from the stagnation point to the edge of the Turing pattern, as expected from the previous study \citep{Pena2001} because the supercriticality $\varepsilon$ was larger closer to the stagnation point. The circular shape of the edge of the Turing pattern was also observed as expected from the theory (Eq.\ref{eq14}), and its radius (about 60 space units in the lowest Pe) was close to the predicted value (46 space units).\\
When the traveling wave coexisted with the Turing pattern, the Turing pattern was formed around the stagnation point, and all the traveling waves started at the interface of the zone of the Turing pattern and the zone of the traveling wave (Fig.\ref{figPorousTuringHopf}a (Multimedia view)). The emergence of a traveling wave can be understood as follows: First, because of the imposed advection, the perturbation does not grow close to the inlet boundary. Also, the spots of the Turing pattern gradually move outward due to the advection (see inside of the black dotted square in Fig.\ref{figPorousTuringHopf}b). When part of the Turing pattern reaches the interface of the two zones, it induces perturbation at the traveling wave zone. This perturbation can grow into a traveling wave given that it is far from the inlet boundary. In this way, the traveling wave starts at the interface of the two zones. The shape of the Turing pattern zone was circular in accordance with the theory (Eq.\ref{eq14}). However, the radius of the circle (about 70 space units) was much smaller than the predicted value from Eq.\ref{eq14} (491 space units). This indicates that the traveling wave appeared even though the theory predicts that the local conditions satisfy Turing instability everywhere. A previous study also showed that the traveling wave may appear even if the instability condition is locally Turing in a reaction diffusion system with heterogeneous reaction parameters \citep{Krause2018}. Our results suggest that the heterogeneity in the diffusion, which corresponds to the dispersion in our study, can also induce the traveling wave under the Turing instability condition.

When there was only Hopf instability in low Pe, the elliptical traveling wave started from close to the stagnation point (Fig.\ref{figLowPe}d, (Multimedia view)). The overall shape of the wave then became like a plane wave as it traveled outward, and the magnitude varied over the plane wave. This is in contrast to the with-diffusion case, where the plane waves started from the unstable manifold and the magnitude of the plane wave was constant in parallel to the unstable manifold. The results of high Pe simulations gave some insight into the underlying mechanisms, which is discussed in the next section.

\subsection{High Pe regime}
\begin{figure}
    \centering
    \includegraphics[width=\textwidth,height=\textheight,keepaspectratio]{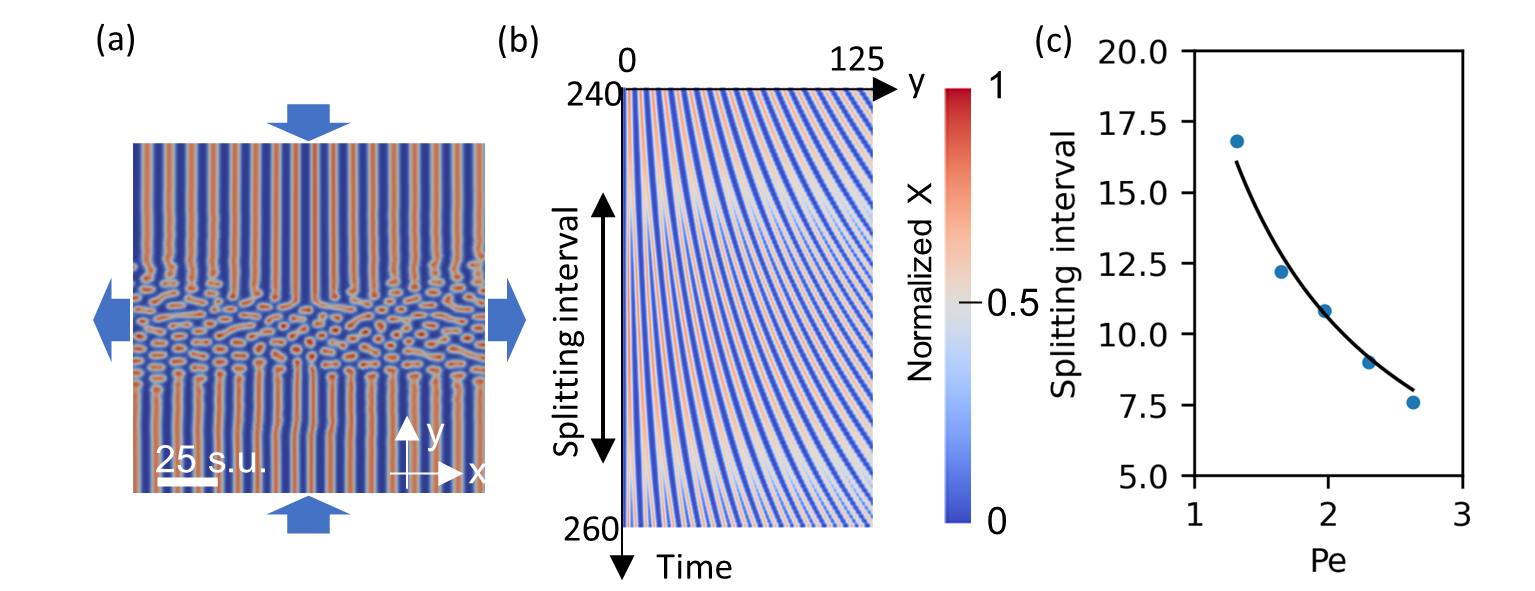}
    \caption{Turing instabilities in with-diffusion cases in high Pe. The blue arrows indicate the flow direction. (a) A snapshot of the normalized concentration field of the species X defined by $(X-min(X)/(max(X)-min(X))$ in high Pe (= 1.64) close to the stagnation point. (Multimedia view) (b) S-T map in high Pe (=1.64), at $y = 50$, $x\in[0,125]$.  The space unit is abbreviated to s.u. Figures (a) and (b) share the same range of concentration. (c) The splitting interval shown in (b) as a function of $Pe$. The black line is the fitted functions, Splitting interval = $a\:Pe^{-1}$, where $a$ is the fitting parameter. The fitted value was $a=21.1118$ with $R^2=0.998$.}
    \label{figNoDispTuring}
\end{figure}

\begin{figure}
    \centering
    \includegraphics[width=\textwidth,height=\textheight,keepaspectratio]{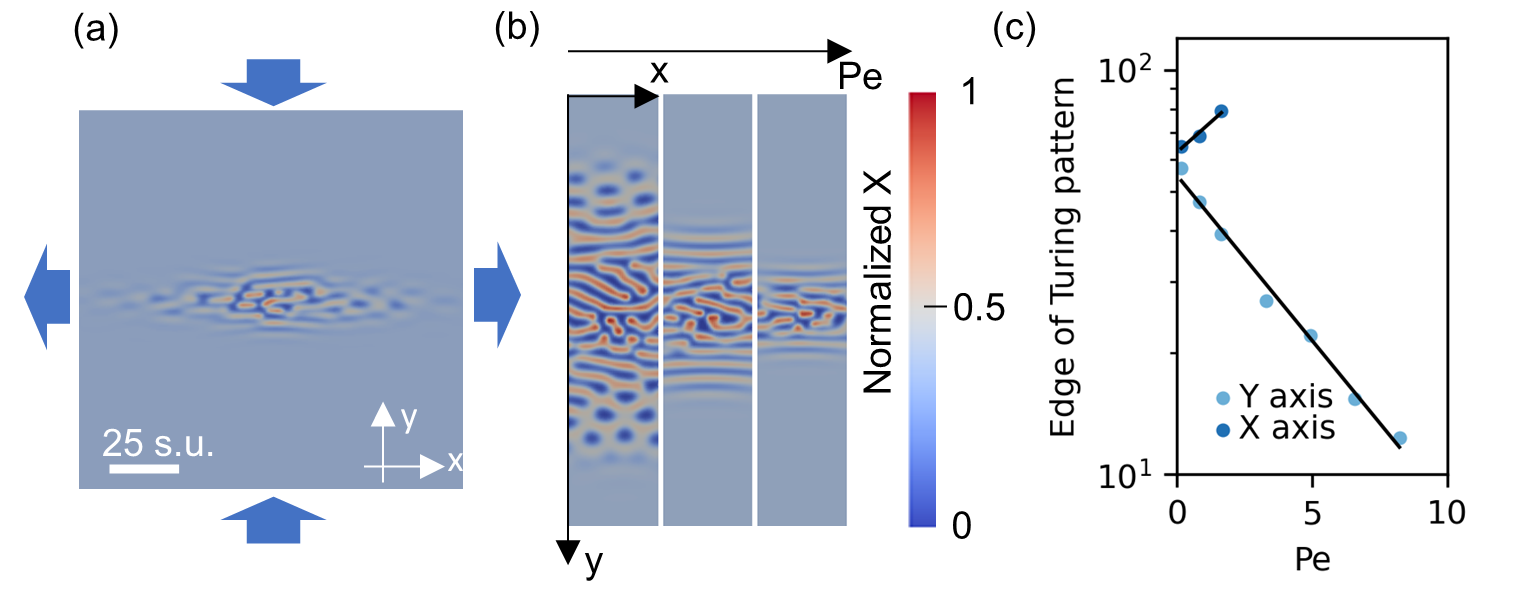}
    \caption{Turing instabilities with-dispersion cases in high Pe. (a) Turing pattern close to the stagnation point in high Pe (= 8.2), showing the normalized concentration of X defined by $(X-min(X)/(max(X)-min(X))$. The blue arrows indicate the flow direction. (Multimedia view). (b) The cropped images in different $Pe\in\{0.82, 3.2, 6.6\}$ in the range $x\in[-12.5,12.5]$, $y \in [0,125]$. The abbreviation s.u. represents a space unit. Figures (a,b) share the same concentration range. (c) The position of the edge of the Turing pattern as a function of $Pe$ on the y-axis (y>0) and x-axis (x>0). The edges along the x-axis at the high $Pe$ range were not clear. Thus, we could not define the edges for these cases. The black lines are the fitted functions, Edge of Turing pattern = $b\: exp(a\:Pe)$, where $a$ and $b$ are the fitting parameters. For the Y axis case, $(a,b)=(-0.1882,4.0051)$ with $R^2=0.990$. For the X axis case, $(a,b)=(0.1370,4.1367)$ with $R^2=0.975$.}
    \label{figPorousTuring}
\end{figure}

We discuss the pattern formation under high Pe in this section. For the Turing instability condition at high Pe in with-diffusion cases, the normal Turing pattern was observed only around the unstable manifold (Fig.\ref{figNoDispTuring}a, (Multimedia view)). In other zones, the advection elongated the Turing patterns in a direction parallel to the stable manifold. As a result, the pattern became stripes. Such a two-dimensional stripe (homoclinic stripe) is known to easily break up into spots in isotropic conditions \citep{Kolokolnikov2018}. A recent study showed that homoclinic stripes can be stabilized when the fast diffusing variable is sufficiently anisotropic \citep{Kolokolnikov2018}. Our results suggest that adding an advection term may also stabilize the homoclinic stripe by introducing anisotropic advective transport. The stripe kept moving parallel to the direction of the unstable manifold (space-time map, Fig.\ref{figNoDispTuring}b). The spacing of the stripe was increasing most of the time except when the lines split intermittently to make the spacing smaller. 
The qualitative explanation of such a splitting stripe is as follows: When the advection is large, the oscillating components are in-phase in the direction of the stable manifold because of $v_x = 0$ and large $v_y$ on the stable manifold. Also, because of the Turing instability, these in-phase components do not oscillate. At the same time, the advection carries these lines toward the outlets, as can be seen in the space-time (S-T) map. The line spacing increases as the lines travel because the fluid velocity is larger at larger x. When the spacing exceeds a certain limit, the lines split to keep the Turing pattern under the given diffusion coefficients. The exception to the stripe pattern is close to the unstable manifold, where the velocity parallel to the stable manifold is very small, and thus the oscillation components are not in-phase. 
Based on the above mechanisms, we calculated the interval time between the splitting of lines (as shown in Fig.\ref{figNoDispTuring}b) as a function of $Pe$. The line spacing increases as $s\propto exp(\gamma t)$ given the saddle flow. The lines split when they reach the critical spacing $s_c$ at time $t_c$. This leads to $t_c \propto \gamma^{-1}ln(s_c)$. Such a proportional relationship between the splitting interval and $Pe^{-1}$ was consistent with the simulation results (Fig.\ref{figNoDispTuring}c), supporting the validity of the above mechanism.\\
When we increased Pe in the Hopf instability condition, we observed the transition from the traveling wave to the stationary pattern (Fig.\ref{figNoDispHopf}a,b, (Multimedia view)). The stationary wave pattern in the Hopf instability condition has also been observed previously, which is known as flow-distributed oscillations (FDO) \citep{Andresen1999,Bamforth2000,Kaern2000,McGraw2005}. In the FDO mechanism, the stationary pattern arises according to the dispersion relation by setting the frequency to zero. The FDO requires (i) the flow to be fast enough and (ii) the boundary concentrations to be stationary. Our boundary condition was the zero gradient. Thus, the concentration at the boundary may change when the traveling wave arrives at the boundary. Therefore, the requirements for FDO matched our simulation conditions only when the flow was fast and the first traveling wave did not reach the boundary (Fig.\ref{figNoDispHopf}a) so that the concentrations at the boundaries were constant. 
To further investigate the role of advection and boundary, we followed the advected fluid element that was placed at the inflow boundary at t = 0. We plotted the y coordinate of the fluid element ($y=125exp(-\gamma t)$) in the S-T map (Fig.\ref{figNoDispHopf}a). The result showed that the traveling wave deflected when it encountered the advected fluid element. Furthermore, the first deflection occurred at the same time in all $Pe$. Thus, given the size of the domain (250) and saddle flow field, the deflection point $y_d$ can be calculated by $y_d=125exp(-\gamma t_d)$, where $t_d$ is the deflection time. We presumed that the position of the y coordinate of the stationary wave $y_w$ would be proportional to $y_d$, which led to $y_w \propto y_d=125exp(-\gamma t_d)\propto 125exp(-Pe)$. Thus, $ln(y_w/125) \propto Pe$. This was consistent with the simulation results (Fig.\ref{figNoDispHopf}c).\\
We further discuss the pattern formations in high Pe with-dispersion cases. In the Turing instability condition, the pattern was elongated parallel to the unstable manifold and shrunk parallel to the stable manifold following the flow direction (Fig.\ref{figPorousTuring}a,b, (Multimedia view)). To quantify such distortion, we plotted the position of the edge of the Turing pattern on the manifolds (Fig.\ref{figPorousTuring}c). The pattern was blurred in high Pe on the unstable manifold. Thus, we limited our analysis to low Pe. The result showed that the edge position is proportional to $exp(-a_1\: Pe)$ and $exp(a_2\: Pe)$ for the edge on the stable (Y axis) and unstable (X axis) manifolds, respectively, with $a_1$ and $a_2$ being positive constants. Such exponential dependency is also observed for the spreading of solute at the stagnation point in saddle flow. When the blob of solute initially occupies the circular domain, the shape of the blob changes following the fluid advection. Specifically, the blob shrinks as $exp(-\gamma t)\propto exp(-Pe\:t)$ along the stable manifold, while it is elongated as $exp(\gamma t)\propto exp(Pe\:t)$ along the stable manifold. The similarity between the shape of the Turing pattern and the spreading of the blob of solute implies that the deformation of the Turing pattern was driven by the advection of the reacted solutes rather than the local instability condition.
This deformation also resulted in the compression of one of the three types of Turing patterns (the outermost reentrant hexagon pattern) for large Pe on the stable manifold (Fig.\ref{figPorousTuring}b), which resulted in the outermost stripe pattern.

\begin{figure}
    \centering
    \includegraphics[width=\textwidth,height=\textheight,keepaspectratio]{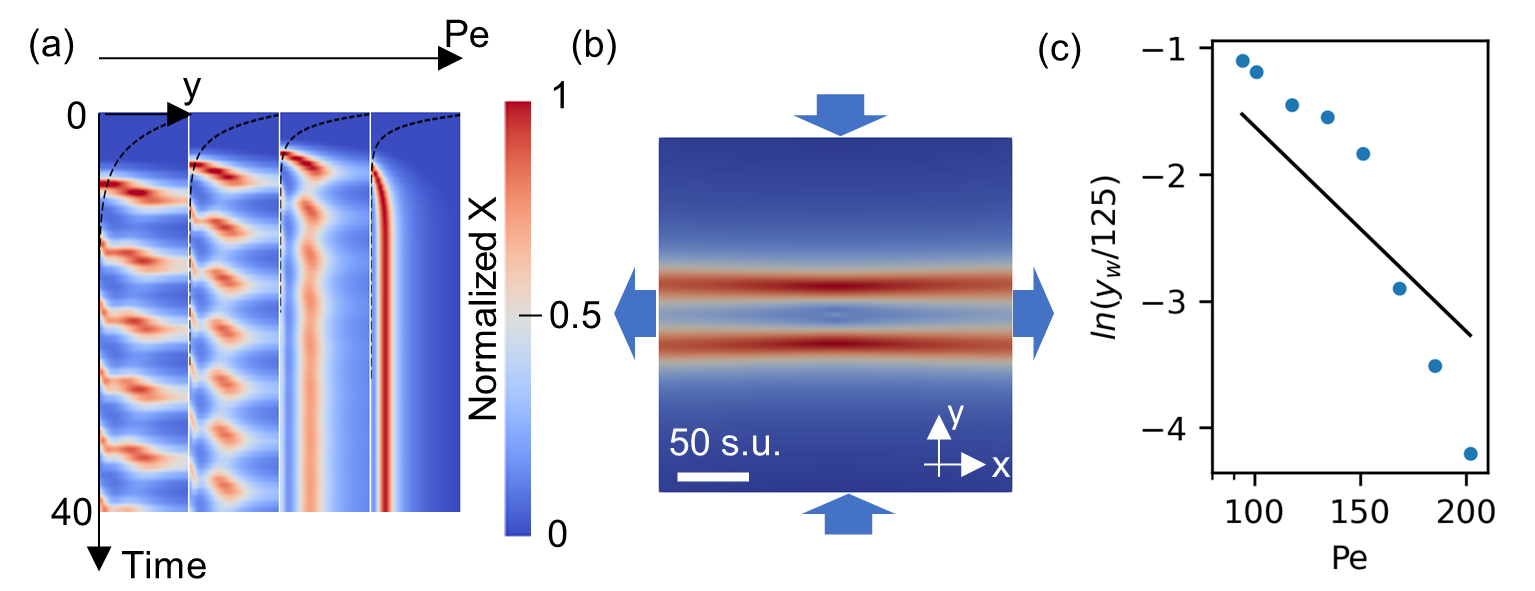}
    \caption{Hopf instability conditions in with-dispersion cases in high Pe. (a) S-T map of the normalized concentration of X defined by $(X-min(X)/(max(X)-min(X))$ in different $Pe\in\{68, 80, 94, 134\}$, at $x = 0$, $y\in[0,125]$. The dotted lines represent the advection of the fluid element that starts at the boundary at time 0. (b) Stationary pattern in high Pe (= 94) (Multimedia view). The blue arrows indicate the flow direction. The abbreviation s.u. represents a space unit. (a) and (b) share the same concentration range. (c) $ln(y_m/125)$ of the stationary wave as a function of $Pe$, where $y_m$ (the position of the wave in y coordinate) is divided by the size of the domain (125). The black line is the fitted functions $ln(y_m/125)=a\: Pe$, where $a$ is the fitting parameter. The fitted value was $a=-0.0162$ with $R^2=0.948$.}
    \label{figPorousHopf}
\end{figure}

\begin{figure}
    \centering
    \includegraphics[width=\textwidth,height=\textheight,keepaspectratio]{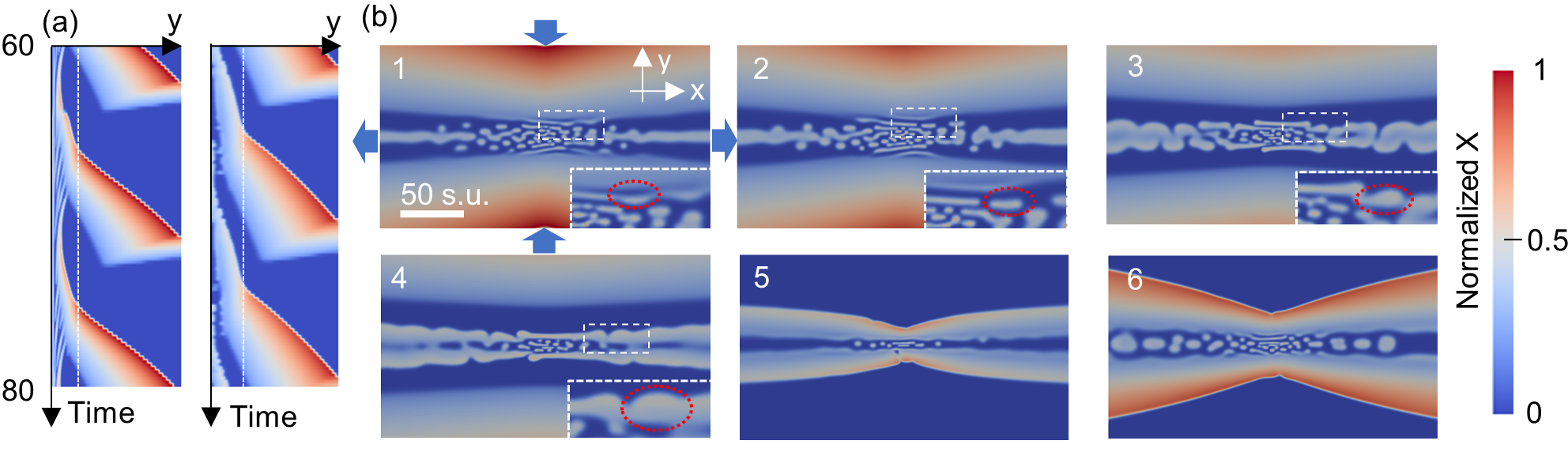}
    \caption{Coexistence of Turing and Hopf instabilities at middle Pe (= 1.32). (a) The S-T map at x = 0 (left), $y\in[0,125]$ and x = 100 (right), $y\in[0,125]$. The left edge of each S-T map corresponds to y = 0. The white dotted line indicates the distance that changes the front speed. (b) Snapshots of the pattern at t = 68.2 (1), t = 68.6 (2), t = 69.6 (3), t = 70.4 (4), t = 73.0 (5), t = 75.0 (6). The zone in the white dotted rectangle is zoomed in at the bottom right for each snapshot. The red-dotted ellipse follows one of the spots in the Turing pattern. The axis and the scale shown in the snapshot 1 are the same for all the snap shot. All the figures share the same concentration range of the normalized concentration of X defined by $(X-min(X)/(max(X)-min(X))$. The blue arrows indicate the flow direction.}
    \label{figPorousTuringHopf_midPe}
\end{figure}

For the condition for coexistance of Turing pattern and traveling wave, the Turing pattern was distorted in the same way as above at the middle Pe (Fig.\ref{figPorousTuringHopf}c, (Multimedia view)), and the traveling wave formed the shape like a ribbon. One of the peculiarities of this pattern was also observed in the space-time map (S-T map) at x = 0,100 (Fig.\ref{figPorousTuringHopf_midPe}a), which shows that the speed of the front was slower closer to the unstable manifold and then became faster when the wave reached a certain distance (indicate by the white dotted line in S-T maps). This is presumably because the instability condition changed from Turing instability to Hopf instability when the front passed a certain distance. Thus, the slow wave would be the moving stripe of the Turing pattern, and the fast wave would be the traveling wave due to the Hopf instability. This interpretation is also consistent with the low Pe case, where the spot in the Turing pattern moved outward with slower speed (Fig.\ref{figPorousTuringHopf}b). The formation of the ribbon shape of the traveling wave (Fig.\ref{figPorousTuringHopf}c) can be explained by following one of the spots of the Turing pattern. We show the snapshots of the wave formation in Fig.\ref{figPorousTuringHopf_midPe}b. First, the Turing pattern was elongated parallel to the unstable manifold, and subsequently, the elongated pattern broke up into small spots (red dotted ellipse in Fig.\ref{figPorousTuringHopf_midPe}b1,2). The isolated spot traveled outward and was blurred at the same time due to the accelerating advection field as well as due to larger dispersion farther from the stagnation point (Fig.\ref{figPorousTuringHopf_midPe}b3). The blurred spot then connected other blurred spots (Fig.\ref{figPorousTuringHopf_midPe}b4). The connected blurred spots became smooth because the advection field $v_y=-\gamma y$ was against the direction of front propagation (i.e., connected spots of the Turing pattern, Fig.\ref{figPorousTuringHopf_midPe}b5). Because the front position farther from the stagnation point (large absolute x coordinate) reached the Hopf instability region earlier (corresponding to the distance indicated by the white line in Fig.\ref{figPorousTuringHopf_midPe}a), only that part of the front became faster while the rest of the front was not yet in Hopf instability region. In this way, the front became like a ribbon shape (Fig.\ref{figPorousTuringHopf_midPe}b6).

\begin{figure}
    \centering
    \includegraphics{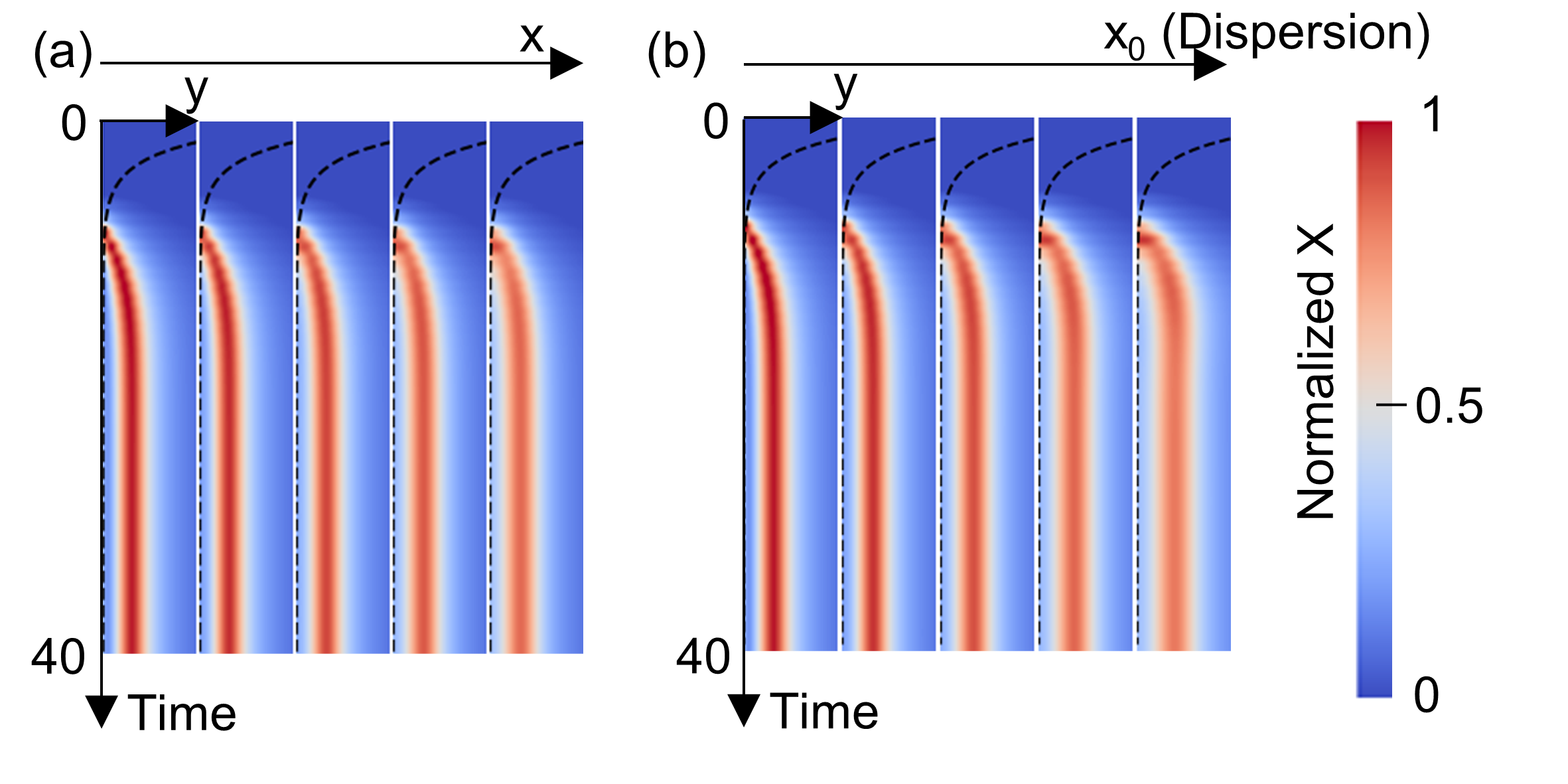}
    \caption{Analysis of Hopf instability condition in with-dispersion cases, showing the S-T map of the normalized concentration of the species X defined by $(X-min(X)/(max(X)-min(X))$. (a) S-T map in $y\in[0,62.5]$ at different $x\in\{0, 25, 50, 75, 100\}$. (b) S-T map in $y\in[0,62.5]$, $x = 0$, with different dispersion fields given by $D_{disp,X}=1+0.444\sqrt{x_0^2+y^2}$, $D_{disp,Y}=0.1+0.444\sqrt{x_0^2+y^2}$, where $x_0\in\{0, 25, 50, 75, 100\}$ and $y$ is the y coordinate (Multimedia view). The right panel corresponds to the highest dipsersion and the left one to the lowest dispersion. The abbreviation s.u. represents a space unit. The dotted lines represent the advection of the fluid element that starts at the boundary at time 0. All the figures share the same concentration range.}
    \label{figPorousHopf_var}
\end{figure}

When we increased Pe only about 5\% from the middle Pe (the above case), the Turing pattern and traveling wave suddenly disappeared, and a stationary pattern arose (Fig.\ref{figPorousTuringHopf}d). The mechanism of this stationary pattern is as follows: In low Pe, traveling waves occurs due to the moving Turing pattern that perturbs the Hopf instability zone. This mechanism also inhibits the formation of FDO at the Hopf instability zone. In turn, in high Pe, the advection becomes large enough to completely suppress the Turing pattern. Thus, the Hopf instability zone is not perturbed, and a stationary pattern arises. \\
For the cases with the Hopf instability condition in the with-dispersion case, the wave stopped within the domain and a stationary pattern arose for high Pe (Fig.\ref{figPorousHopf}a,b,(Multimedia view)). We analyzed the position of the stationary waves as in the with-diffusion case, where we plotted $ln(y_x/125)$ (the position of the wave on the y-axis is $y_w$, which is divided by the half size of the domain) (Fig.\ref{figPorousHopf}c). In contrast to the with-diffusion case, we did not observe $ln(y_x/125)\propto Pe$ in the entire Pe range, indicating that the heterogeneous dispersion field also plays a role in determining the position of the stationary waves.\\
The main qualitative difference between the with-diffusion case and the with-dispersion case under Hopf instability condition was the shape of the frozen wave. In the with-dispersion cases, the frozen wave was elliptical in contrast to the plane waves in the with-diffusion cases. This suggests that the wave starting from the stagnation point not only stopped against the inlet flow (parallel to the stable manifold) but also stopped even where the flow direction followed the direction of the traveling wave (parallel to the unstable manifold). This is also in contrast to the case of an autocatalytic reaction, where the advection is only a one-way barrier to the wave propagation \citep{Mahoney2012,Mahoney2015,Mitchell2012,Doan2018}. The S-T map at different x positions (Fig.\ref{figPorousHopf_var}a) showed that the stationary pattern formed at the same timing everywhere. This indicates that the formation of a stationary wave pattern may be understood without considering the advection and dispersion in the x direction. To check this hypothesis, we have run additional simulations to investigate how pattern formation changes when dispersion fields vary only in the y direction. From Eq.\ref{eq10}, the dispersion field was given by $D_{disp,X}=1+0.444\sqrt{x_0^2+y^2}$, $D_{disp,Y}=0.1+0.444\sqrt{x_0^2+y^2}$, where $x_0$ varied in different scenarios as $x_0\in\{0, 25, 50, 75, 100\}$ and $y$ was the y coordinate (see Eq.\ref{eq9},\ref{eq10}). In this way, the dispersion field varied only in the y direction. Note that the dispersion field of Fig.\ref{figPorousHopf}b was given by $D_{disp,X}=1+0.444\sqrt{x^2+y^2}$, $D_{disp,Y}=0.1+0.444\sqrt{x^2+y^2}$, which led to the variation of the dispersion in both x and y directions. Thus, each scenario with a fixed value for $x_0$ replicated the variation of dispersion field at each $x$ position in the simulation of Fig.\ref{figPorousHopf}b. The advection field and concentrations of $A_0$ and $B_0$ were the same as the original Hopf instability case shown in Fig\ref{figPorousHopf}b ($(A_0,B_0,\gamma) = (1,2.2,0.592)$). The results showed frozen plane waves similar to those observed in the with-diffusion case shown in Fig.\ref{figNoDispHopf}b (Fig.\ref{figPorousHopf_var}b, (Multimedia view) for the case of $x_0=50$). More importantly, the S-T map at x = 0 and $y\in[0,62.5]$ (Fig.\ref{figPorousHopf_var}b) showed that each scenario successfully replicated the frozen wave formation at each x position in the original case (Fig.\ref{figPorousHopf_var}a), where larger x resulted in waves with smaller magnitude and larger wave length. A little larger magnitude in the original case at large x (e.g., right panel in Fig.\ref{figPorousHopf_var}a) compared to the corresponding scenario (e.g., right panel in Fig.\ref{figPorousHopf_var}b) shows that the upstream had a small influence on the downstream, but the effect was minor. The above analysis clarified that the frozen elliptical wave was formed by the accumulation of the frozen plane waves that were formed at different distances from the unstable manifold.\\

\section{Summary and conclusion}
We investigated pattern formation in porous media under saddle flow only with diffusion and both with diffusion and dispersion. With-diffusion cases represented the pore scale flow field, while with-dispersion cases represented the continuum-scale transport in porous media. The theory predicted four distinct scenarios when dispersion controls solute spreading: (I) homogeneous steady state; (II) local Turing instability around the stagnation point; (III) Turing instability around the stagnation point and Hopf instability away from the stagnation point and (IV) Hopf instability. Next, we have run numerical simulations corresponding to each scenario in the with-diffusion and with-dispersion cases. In the with-diffusion cases,  peculiar patterns were observed for high Pe including the elongated Turing pattern parallel to the stable manifold that continuously divided and the frozen traveling waves that were parallel to the unstable manifold. \\
The simulations of with-dispersion cases in low Pe showed that, for the Turing instability condition, the localized Turing pattern appeared around the stagnation point, as expected from the theory. When the Turing instability coexisted with the Hopf instability, the size of the localized Turing pattern around the stagnation point significantly differed from the theoretical prediction, suggesting that a non-local instability condition should be considered. For the Hopf instability condition, unlike the plane wave in with-diffusion cases, elliptical traveling waves occurred that started from the stagnation point.\\
When we increased Pe in with-dispersion cases, the wave pattern changed qualitatively due to dominant advection over dispersion. For the Turing instability condition, the localized Turing pattern was elongated in the direction of the unstable manifold and compressed in the direction of the stable manifold. For the coexistence of the Turing and Hopf instability conditions at middle Pe, the traveling wave accelerated when it crossed the border between the Turing instability condition and the Hopf instability condition. Also, a "ribbon" shaped front appeared due to the travel of spots of the Turing pattern. In high Pe, the stationary pattern appeared when the Turing pattern was annihilated. For the Hopf instability condition, the high Pe induced the flow distributed oscillator. Unlike the flow distributed oscillator in with-diffusion cases, the stationary wave was elliptical around the stagnation point instead of plane, even where the direction of the wave was not opposing the direction of flow. Further analysis revealed that this elliptical stationary wave was actually the accumulation of the plane stationary waves that were formed at different distances from the unstable manifold.\\
Our results give the first classification of the pattern formation of the oscillating reaction in porous media at the pore-scale and at the continuum-scale with fluid deformation. We highlight that the pattern was controlled by the stagnation point, stable and unstable manifolds, and Pe. If experimentally achieved, such localized pattern formations would benefit engineering applications of oscillating reactions such as processing and storing information through chemical computation processors\citep{Ryzhkov2021,Parrilla-Gutierrez2020,Wang2016} and mimicking the localized pattern formation in biological systems. Furthermore, the predator-prey system\citep{Goldbeter2017,Goldbeter2018} in porous media in nature potentially shows some of the patterns in our study. Such patterns would be understood by applying the interpretations shown in our study.\\
One of the natural extensions of our study is to introduce one more dimension to simulate three-dimensional pattern formation in porous media. In this case, the saddle flow should be categorized into two patterns: converging saddle and diverging saddle, where the former flow has two directions for converging streamlines and the latter flow has two directions for diverging streamlines \citep{Bresciani2019}. Thus, such classification will be required to study three-dimensional pattern formation. As for the patterns, we observed circular or elliptical Turing patterns around the stagnation point in two-dimensional space. In three-dimensional space, it would become a sphere, oblate, or prolate depending on the flow rate and diverging or converging saddle flow. Furthermore, the manifold is line in two-dimensional flow, whereas it becomes sheet in three-dimensional flow \citep{Metcalfe2022}. Therefore, the plane line waves observed in our study would become flat sheet waves in three-dimensional space. The extension to three-dimensional simulations requires further optimization of simulation methods to investigate efficiently. Thus, we leave it for future studies.\\

\begin{acknowledgments}
We acknowledge Prof. Anne De Wit and Dr. Dario M. Escala in Nonlinear Physical Chemistry Unit, Universit\'e libre de Bruxelles (ULB) for insightful discussion.
\end{acknowledgments}

\appendix



\begin{thebibliography}{76}%
\makeatletter
\providecommand \@ifxundefined [1]{%
 \@ifx{#1\undefined}
}%
\providecommand \@ifnum [1]{%
 \ifnum #1\expandafter \@firstoftwo
 \else \expandafter \@secondoftwo
 \fi
}%
\providecommand \@ifx [1]{%
 \ifx #1\expandafter \@firstoftwo
 \else \expandafter \@secondoftwo
 \fi
}%
\providecommand \natexlab [1]{#1}%
\providecommand \enquote  [1]{``#1''}%
\providecommand \bibnamefont  [1]{#1}%
\providecommand \bibfnamefont [1]{#1}%
\providecommand \citenamefont [1]{#1}%
\providecommand \href@noop [0]{\@secondoftwo}%
\providecommand \href [0]{\begingroup \@sanitize@url \@href}%
\providecommand \@href[1]{\@@startlink{#1}\@@href}%
\providecommand \@@href[1]{\endgroup#1\@@endlink}%
\providecommand \@sanitize@url [0]{\catcode `\\12\catcode `\$12\catcode
  `\&12\catcode `\#12\catcode `\^12\catcode `\_12\catcode `\%12\relax}%
\providecommand \@@startlink[1]{}%
\providecommand \@@endlink[0]{}%
\providecommand \url  [0]{\begingroup\@sanitize@url \@url }%
\providecommand \@url [1]{\endgroup\@href {#1}{\urlprefix }}%
\providecommand \urlprefix  [0]{URL }%
\providecommand \Eprint [0]{\href }%
\providecommand \doibase [0]{https://doi.org/}%
\providecommand \selectlanguage [0]{\@gobble}%
\providecommand \bibinfo  [0]{\@secondoftwo}%
\providecommand \bibfield  [0]{\@secondoftwo}%
\providecommand \translation [1]{[#1]}%
\providecommand \BibitemOpen [0]{}%
\providecommand \bibitemStop [0]{}%
\providecommand \bibitemNoStop [0]{.\EOS\space}%
\providecommand \EOS [0]{\spacefactor3000\relax}%
\providecommand \BibitemShut  [1]{\csname bibitem#1\endcsname}%
\let\auto@bib@innerbib\@empty
\bibitem [{\citenamefont {Cross}\ and\ \citenamefont
  {Hohenberg}(1993)}]{Cross1993}%
  \BibitemOpen
  \bibfield  {author} {\bibinfo {author} {\bibfnamefont {M.~C.}\ \bibnamefont
  {Cross}}\ and\ \bibinfo {author} {\bibfnamefont {P.~C.}\ \bibnamefont
  {Hohenberg}},\ }\href {https://doi.org/10.1103/RevModPhys.65.851} {\bibfield
  {journal} {\bibinfo  {journal} {Reviews of Modern Physics}\ }\textbf
  {\bibinfo {volume} {65}},\ \bibinfo {pages} {851} (\bibinfo {year}
  {1993})}\BibitemShut {NoStop}%
\bibitem [{\citenamefont {Haudin}\ \emph {et~al.}(2009)\citenamefont {Haudin},
  \citenamefont {El{\'{i}}as}, \citenamefont {Rojas}, \citenamefont
  {Bortolozzo}, \citenamefont {Clerc},\ and\ \citenamefont
  {Residori}}]{Haudin2009}%
  \BibitemOpen
  \bibfield  {author} {\bibinfo {author} {\bibfnamefont {F.}~\bibnamefont
  {Haudin}}, \bibinfo {author} {\bibfnamefont {R.~G.}\ \bibnamefont
  {El{\'{i}}as}}, \bibinfo {author} {\bibfnamefont {R.~G.}\ \bibnamefont
  {Rojas}}, \bibinfo {author} {\bibfnamefont {U.}~\bibnamefont {Bortolozzo}},
  \bibinfo {author} {\bibfnamefont {M.~G.}\ \bibnamefont {Clerc}},\ and\
  \bibinfo {author} {\bibfnamefont {S.}~\bibnamefont {Residori}},\ }\href
  {https://doi.org/10.1103/PhysRevLett.103.128003} {\bibfield  {journal}
  {\bibinfo  {journal} {Physical Review Letters}\ }\textbf {\bibinfo {volume}
  {103}},\ \bibinfo {pages} {1} (\bibinfo {year} {2009})}\BibitemShut {NoStop}%
\bibitem [{\citenamefont {Van~Saarloos}(2003)}]{VanSaarloos2003}%
  \BibitemOpen
  \bibfield  {author} {\bibinfo {author} {\bibfnamefont {W.}~\bibnamefont
  {Van~Saarloos}},\ }\href {https://doi.org/10.1016/j.physrep.2003.08.001}
  {\bibfield  {journal} {\bibinfo  {journal} {Physics Reports}\ }\textbf
  {\bibinfo {volume} {386}},\ \bibinfo {pages} {29} (\bibinfo {year}
  {2003})}\BibitemShut {NoStop}%
\bibitem [{\citenamefont {Bressloff}\ and\ \citenamefont
  {Newby}(2013)}]{Bressloff2013}%
  \BibitemOpen
  \bibfield  {author} {\bibinfo {author} {\bibfnamefont {P.~C.}\ \bibnamefont
  {Bressloff}}\ and\ \bibinfo {author} {\bibfnamefont {J.~M.}\ \bibnamefont
  {Newby}},\ }\href {https://doi.org/10.1103/RevModPhys.85.135} {\bibfield
  {journal} {\bibinfo  {journal} {Reviews of Modern Physics}\ }\textbf
  {\bibinfo {volume} {85}},\ \bibinfo {pages} {135} (\bibinfo {year}
  {2013})}\BibitemShut {NoStop}%
\bibitem [{\citenamefont {Beta}\ and\ \citenamefont {Kruse}(2017)}]{Beta2017}%
  \BibitemOpen
  \bibfield  {author} {\bibinfo {author} {\bibfnamefont {C.}~\bibnamefont
  {Beta}}\ and\ \bibinfo {author} {\bibfnamefont {K.}~\bibnamefont {Kruse}},\
  }\href {https://doi.org/10.1146/annurev-conmatphys-031016-025210} {\bibfield
  {journal} {\bibinfo  {journal} {Annual Review of Condensed Matter Physics}\
  }\textbf {\bibinfo {volume} {8}},\ \bibinfo {pages} {239} (\bibinfo {year}
  {2017})}\BibitemShut {NoStop}%
\bibitem [{\citenamefont {Sagu{\'{e}}s}\ and\ \citenamefont
  {Epstein}(2003)}]{Sagues2003}%
  \BibitemOpen
  \bibfield  {author} {\bibinfo {author} {\bibfnamefont {F.}~\bibnamefont
  {Sagu{\'{e}}s}}\ and\ \bibinfo {author} {\bibfnamefont {I.~R.}\ \bibnamefont
  {Epstein}},\ }\href {https://doi.org/10.1039/b210932h} {\bibfield  {journal}
  {\bibinfo  {journal} {Dalton Transactions}\ ,\ \bibinfo {pages} {1201}}
  (\bibinfo {year} {2003})}\BibitemShut {NoStop}%
\bibitem [{\citenamefont {Rotermund}(2009)}]{Rotermund2009}%
  \BibitemOpen
  \bibfield  {author} {\bibinfo {author} {\bibfnamefont {H.~H.}\ \bibnamefont
  {Rotermund}},\ }\href {https://doi.org/10.1016/j.susc.2008.11.048} {\bibfield
   {journal} {\bibinfo  {journal} {Surface Science}\ }\textbf {\bibinfo
  {volume} {603}},\ \bibinfo {pages} {1662} (\bibinfo {year}
  {2009})}\BibitemShut {NoStop}%
\bibitem [{\citenamefont {Samuelson}\ \emph {et~al.}(2019)\citenamefont
  {Samuelson}, \citenamefont {Singer}, \citenamefont {Weinburd},\ and\
  \citenamefont {Scheel}}]{Samuelson2019}%
  \BibitemOpen
  \bibfield  {author} {\bibinfo {author} {\bibfnamefont {R.}~\bibnamefont
  {Samuelson}}, \bibinfo {author} {\bibfnamefont {Z.}~\bibnamefont {Singer}},
  \bibinfo {author} {\bibfnamefont {J.}~\bibnamefont {Weinburd}},\ and\
  \bibinfo {author} {\bibfnamefont {A.}~\bibnamefont {Scheel}},\ }\href
  {https://doi.org/10.1007/s00332-018-9486-6} {\bibfield  {journal} {\bibinfo
  {journal} {Journal of Nonlinear Science}\ }\textbf {\bibinfo {volume} {29}},\
  \bibinfo {pages} {255} (\bibinfo {year} {2019})}\BibitemShut {NoStop}%
\bibitem [{\citenamefont {Martin}(2003)}]{Martin2003a}%
  \BibitemOpen
  \bibfield  {author} {\bibinfo {author} {\bibfnamefont {A.}~\bibnamefont
  {Martin}},\ }\href {https://doi.org/10.1016/s0079-6611(03)00085-5} {\bibfield
   {journal} {\bibinfo  {journal} {Progress in Oceanography}\ }\textbf
  {\bibinfo {volume} {57}},\ \bibinfo {pages} {125} (\bibinfo {year}
  {2003})}\BibitemShut {NoStop}%
\bibitem [{\citenamefont {Nicolis}\ and\ \citenamefont
  {Prigogine}(1977)}]{nicolis1977}%
  \BibitemOpen
  \bibfield  {author} {\bibinfo {author} {\bibfnamefont {G.}~\bibnamefont
  {Nicolis}}\ and\ \bibinfo {author} {\bibfnamefont {I.}~\bibnamefont
  {Prigogine}},\ }\href@noop {} {\emph {\bibinfo {title} {{Self-Organization in
  Nonequilibrium Systems: From Dissipative Structures to Order Through
  Fluctuations}}}},\ \bibinfo {edition} {new york}\ ed.\ (\bibinfo  {publisher}
  {Wiley},\ \bibinfo {year} {1977})\BibitemShut {NoStop}%
\bibitem [{\citenamefont {Doan}\ \emph {et~al.}(2018)\citenamefont {Doan},
  \citenamefont {Simons}, \citenamefont {Lilienthal}, \citenamefont {Solomon},\
  and\ \citenamefont {Mitchell}}]{Doan2018}%
  \BibitemOpen
  \bibfield  {author} {\bibinfo {author} {\bibfnamefont {M.}~\bibnamefont
  {Doan}}, \bibinfo {author} {\bibfnamefont {J.~J.}\ \bibnamefont {Simons}},
  \bibinfo {author} {\bibfnamefont {K.}~\bibnamefont {Lilienthal}}, \bibinfo
  {author} {\bibfnamefont {T.}~\bibnamefont {Solomon}},\ and\ \bibinfo {author}
  {\bibfnamefont {K.~A.}\ \bibnamefont {Mitchell}},\ }\href
  {https://doi.org/10.1103/PhysRevE.97.033111} {\bibfield  {journal} {\bibinfo
  {journal} {Physical Review E}\ }\textbf {\bibinfo {volume} {97}},\ \bibinfo
  {pages} {033111} (\bibinfo {year} {2018})}\BibitemShut {NoStop}%
\bibitem [{\citenamefont {Nevins}\ and\ \citenamefont
  {Kelley}(2016)}]{Nevins2016}%
  \BibitemOpen
  \bibfield  {author} {\bibinfo {author} {\bibfnamefont {T.~D.}\ \bibnamefont
  {Nevins}}\ and\ \bibinfo {author} {\bibfnamefont {D.~H.}\ \bibnamefont
  {Kelley}},\ }\href {https://doi.org/10.1103/PhysRevLett.117.164502}
  {\bibfield  {journal} {\bibinfo  {journal} {Physical Review Letters}\
  }\textbf {\bibinfo {volume} {117}},\ \bibinfo {pages} {1} (\bibinfo {year}
  {2016})}\BibitemShut {NoStop}%
\bibitem [{\citenamefont {Nugent}\ \emph {et~al.}(2004)\citenamefont {Nugent},
  \citenamefont {Quarles},\ and\ \citenamefont {Solomon}}]{Nugent2004}%
  \BibitemOpen
  \bibfield  {author} {\bibinfo {author} {\bibfnamefont {C.~R.}\ \bibnamefont
  {Nugent}}, \bibinfo {author} {\bibfnamefont {W.~M.}\ \bibnamefont
  {Quarles}},\ and\ \bibinfo {author} {\bibfnamefont {T.~H.}\ \bibnamefont
  {Solomon}},\ }\href {https://doi.org/10.1103/PhysRevLett.93.218301}
  {\bibfield  {journal} {\bibinfo  {journal} {Physical Review Letters}\
  }\textbf {\bibinfo {volume} {93}},\ \bibinfo {pages} {218301} (\bibinfo
  {year} {2004})}\BibitemShut {NoStop}%
\bibitem [{\citenamefont {Vasquez}(2004)}]{Vasquez2004}%
  \BibitemOpen
  \bibfield  {author} {\bibinfo {author} {\bibfnamefont {D.~A.}\ \bibnamefont
  {Vasquez}},\ }\href {https://doi.org/10.1103/PhysRevLett.93.104501}
  {\bibfield  {journal} {\bibinfo  {journal} {Physical Review Letters}\
  }\textbf {\bibinfo {volume} {93}},\ \bibinfo {pages} {3} (\bibinfo {year}
  {2004})}\BibitemShut {NoStop}%
\bibitem [{\citenamefont {Neufeld}(2001)}]{Neufeld2001}%
  \BibitemOpen
  \bibfield  {author} {\bibinfo {author} {\bibfnamefont {Z.}~\bibnamefont
  {Neufeld}},\ }\href {https://doi.org/10.1103/PhysRevLett.87.108301}
  {\bibfield  {journal} {\bibinfo  {journal} {Physical Review Letters}\
  }\textbf {\bibinfo {volume} {87}},\ \bibinfo {pages} {108301} (\bibinfo
  {year} {2001})}\BibitemShut {NoStop}%
\bibitem [{\citenamefont {Escala}\ \emph {et~al.}(2014)\citenamefont {Escala},
  \citenamefont {Budroni}, \citenamefont {Carballido-Landeira}, \citenamefont
  {De~Wit},\ and\ \citenamefont {Mu{\~{n}}uzuri}}]{Escala2014}%
  \BibitemOpen
  \bibfield  {author} {\bibinfo {author} {\bibfnamefont {D.~M.}\ \bibnamefont
  {Escala}}, \bibinfo {author} {\bibfnamefont {M.~A.}\ \bibnamefont {Budroni}},
  \bibinfo {author} {\bibfnamefont {J.}~\bibnamefont {Carballido-Landeira}},
  \bibinfo {author} {\bibfnamefont {A.}~\bibnamefont {De~Wit}},\ and\ \bibinfo
  {author} {\bibfnamefont {A.~P.}\ \bibnamefont {Mu{\~{n}}uzuri}},\ }\href
  {https://doi.org/10.1021/jz402625z} {\bibfield  {journal} {\bibinfo
  {journal} {Journal of Physical Chemistry Letters}\ }\textbf {\bibinfo
  {volume} {5}},\ \bibinfo {pages} {413} (\bibinfo {year} {2014})}\BibitemShut
  {NoStop}%
\bibitem [{\citenamefont {Schwartz}\ and\ \citenamefont
  {Solomon}(2008)}]{Schwartz2008}%
  \BibitemOpen
  \bibfield  {author} {\bibinfo {author} {\bibfnamefont {M.~E.}\ \bibnamefont
  {Schwartz}}\ and\ \bibinfo {author} {\bibfnamefont {T.~H.}\ \bibnamefont
  {Solomon}},\ }\href {https://doi.org/10.1103/PhysRevLett.100.028302}
  {\bibfield  {journal} {\bibinfo  {journal} {Physical Review Letters}\
  }\textbf {\bibinfo {volume} {100}},\ \bibinfo {pages} {028302} (\bibinfo
  {year} {2008})}\BibitemShut {NoStop}%
\bibitem [{\citenamefont {Manz}\ \emph {et~al.}(2004)\citenamefont {Manz},
  \citenamefont {Hamik},\ and\ \citenamefont {Steinbock}}]{Manz2004}%
  \BibitemOpen
  \bibfield  {author} {\bibinfo {author} {\bibfnamefont {N.}~\bibnamefont
  {Manz}}, \bibinfo {author} {\bibfnamefont {C.~T.}\ \bibnamefont {Hamik}},\
  and\ \bibinfo {author} {\bibfnamefont {O.}~\bibnamefont {Steinbock}},\ }\href
  {https://doi.org/10.1103/PhysRevLett.92.248301} {\bibfield  {journal}
  {\bibinfo  {journal} {Physical Review Letters}\ }\textbf {\bibinfo {volume}
  {92}},\ \bibinfo {pages} {1} (\bibinfo {year} {2004})}\BibitemShut {NoStop}%
\bibitem [{\citenamefont {Golovin}\ \emph {et~al.}(2008)\citenamefont
  {Golovin}, \citenamefont {Matkowsky},\ and\ \citenamefont
  {Volpert}}]{Golovin2008}%
  \BibitemOpen
  \bibfield  {author} {\bibinfo {author} {\bibfnamefont {A.~A.}\ \bibnamefont
  {Golovin}}, \bibinfo {author} {\bibfnamefont {B.~J.}\ \bibnamefont
  {Matkowsky}},\ and\ \bibinfo {author} {\bibfnamefont {V.~A.}\ \bibnamefont
  {Volpert}},\ }\href {https://doi.org/10.1137/070703454} {\bibfield  {journal}
  {\bibinfo  {journal} {SIAM Journal on Applied Mathematics}\ }\textbf
  {\bibinfo {volume} {69}},\ \bibinfo {pages} {251} (\bibinfo {year}
  {2008})}\BibitemShut {NoStop}%
\bibitem [{\citenamefont {Berenstein}\ and\ \citenamefont
  {Beta}(2013)}]{Berenstein2013}%
  \BibitemOpen
  \bibfield  {author} {\bibinfo {author} {\bibfnamefont {I.}~\bibnamefont
  {Berenstein}}\ and\ \bibinfo {author} {\bibfnamefont {C.}~\bibnamefont
  {Beta}},\ }\href {https://doi.org/10.1063/1.4816937} {\bibfield  {journal}
  {\bibinfo  {journal} {Chaos: An Interdisciplinary Journal of Nonlinear
  Science}\ }\textbf {\bibinfo {volume} {23}},\ \bibinfo {pages} {033119}
  (\bibinfo {year} {2013})}\BibitemShut {NoStop}%
\bibitem [{\citenamefont {Paoletti}\ \emph {et~al.}(2006)\citenamefont
  {Paoletti}, \citenamefont {Nugent},\ and\ \citenamefont
  {Solomon}}]{Paoletti2006}%
  \BibitemOpen
  \bibfield  {author} {\bibinfo {author} {\bibfnamefont {M.~S.}\ \bibnamefont
  {Paoletti}}, \bibinfo {author} {\bibfnamefont {C.~R.}\ \bibnamefont
  {Nugent}},\ and\ \bibinfo {author} {\bibfnamefont {T.~H.}\ \bibnamefont
  {Solomon}},\ }\href {https://doi.org/10.1103/PhysRevLett.96.124101}
  {\bibfield  {journal} {\bibinfo  {journal} {Physical Review Letters}\
  }\textbf {\bibinfo {volume} {96}},\ \bibinfo {pages} {124101} (\bibinfo
  {year} {2006})}\BibitemShut {NoStop}%
\bibitem [{\citenamefont {Ginn}\ \emph {et~al.}(2004)\citenamefont {Ginn},
  \citenamefont {Steinbock}, \citenamefont {Kahveci},\ and\ \citenamefont
  {Steinbock}}]{Ginn2004a}%
  \BibitemOpen
  \bibfield  {author} {\bibinfo {author} {\bibfnamefont {B.~T.}\ \bibnamefont
  {Ginn}}, \bibinfo {author} {\bibfnamefont {B.}~\bibnamefont {Steinbock}},
  \bibinfo {author} {\bibfnamefont {M.}~\bibnamefont {Kahveci}},\ and\ \bibinfo
  {author} {\bibfnamefont {O.}~\bibnamefont {Steinbock}},\ }\href
  {https://doi.org/10.1021/jp0358883} {\bibfield  {journal} {\bibinfo
  {journal} {Journal of Physical Chemistry A}\ }\textbf {\bibinfo {volume}
  {108}},\ \bibinfo {pages} {1325} (\bibinfo {year} {2004})}\BibitemShut
  {NoStop}%
\bibitem [{\citenamefont {Koz{\'{a}}k}\ \emph {et~al.}(2019)\citenamefont
  {Koz{\'{a}}k}, \citenamefont {Gaffney},\ and\ \citenamefont
  {Klika}}]{Kozak2019}%
  \BibitemOpen
  \bibfield  {author} {\bibinfo {author} {\bibfnamefont {M.}~\bibnamefont
  {Koz{\'{a}}k}}, \bibinfo {author} {\bibfnamefont {E.~A.}\ \bibnamefont
  {Gaffney}},\ and\ \bibinfo {author} {\bibfnamefont {V.}~\bibnamefont
  {Klika}},\ }\href {https://doi.org/10.1103/PhysRevE.100.042220} {\bibfield
  {journal} {\bibinfo  {journal} {Physical Review E}\ }\textbf {\bibinfo
  {volume} {100}},\ \bibinfo {pages} {042220} (\bibinfo {year}
  {2019})}\BibitemShut {NoStop}%
\bibitem [{\citenamefont {Wang}\ \emph {et~al.}(2017)\citenamefont {Wang},
  \citenamefont {Tithof}, \citenamefont {Nevins}, \citenamefont {Col{\'{o}}n},\
  and\ \citenamefont {Kelley}}]{Wang2017}%
  \BibitemOpen
  \bibfield  {author} {\bibinfo {author} {\bibfnamefont {J.}~\bibnamefont
  {Wang}}, \bibinfo {author} {\bibfnamefont {J.}~\bibnamefont {Tithof}},
  \bibinfo {author} {\bibfnamefont {T.~D.}\ \bibnamefont {Nevins}}, \bibinfo
  {author} {\bibfnamefont {R.~O.}\ \bibnamefont {Col{\'{o}}n}},\ and\ \bibinfo
  {author} {\bibfnamefont {D.~H.}\ \bibnamefont {Kelley}},\ }\href
  {https://doi.org/10.1063/1.5004649} {\bibfield  {journal} {\bibinfo
  {journal} {Chaos: An Interdisciplinary Journal of Nonlinear Science}\
  }\textbf {\bibinfo {volume} {27}},\ \bibinfo {pages} {123109} (\bibinfo
  {year} {2017})}\BibitemShut {NoStop}%
\bibitem [{\citenamefont {Klika}\ \emph {et~al.}(2018)\citenamefont {Klika},
  \citenamefont {Koz{\'{a}}k},\ and\ \citenamefont {Gaffney}}]{klika2018}%
  \BibitemOpen
  \bibfield  {author} {\bibinfo {author} {\bibfnamefont {V.}~\bibnamefont
  {Klika}}, \bibinfo {author} {\bibfnamefont {M.}~\bibnamefont {Koz{\'{a}}k}},\
  and\ \bibinfo {author} {\bibfnamefont {E.~A.}\ \bibnamefont {Gaffney}},\
  }\href {https://doi.org/10.1137/17M1138571} {\bibfield  {journal} {\bibinfo
  {journal} {SIAM Journal on Applied Mathematics}\ }\textbf {\bibinfo {volume}
  {78}},\ \bibinfo {pages} {2298} (\bibinfo {year} {2018})}\BibitemShut
  {NoStop}%
\bibitem [{\citenamefont {Ryzhkov}\ \emph {et~al.}(2021)\citenamefont
  {Ryzhkov}, \citenamefont {Nikolaev}, \citenamefont {Ivanov},\ and\
  \citenamefont {Skorb}}]{Ryzhkov2021}%
  \BibitemOpen
  \bibfield  {author} {\bibinfo {author} {\bibfnamefont {N.~V.}\ \bibnamefont
  {Ryzhkov}}, \bibinfo {author} {\bibfnamefont {K.~G.}\ \bibnamefont
  {Nikolaev}}, \bibinfo {author} {\bibfnamefont {A.~S.}\ \bibnamefont
  {Ivanov}},\ and\ \bibinfo {author} {\bibfnamefont {E.~V.}\ \bibnamefont
  {Skorb}},\ }\href {https://doi.org/10.1146/annurev-chembioeng-122120-023514}
  {\bibfield  {journal} {\bibinfo  {journal} {Annual Review of Chemical and
  Biomolecular Engineering}\ }\textbf {\bibinfo {volume} {12}},\ \bibinfo
  {pages} {63} (\bibinfo {year} {2021})}\BibitemShut {NoStop}%
\bibitem [{\citenamefont {Parrilla-Gutierrez}\ \emph
  {et~al.}(2020)\citenamefont {Parrilla-Gutierrez}, \citenamefont {Sharma},
  \citenamefont {Tsuda}, \citenamefont {Cooper}, \citenamefont
  {Aragon-Camarasa}, \citenamefont {Donkers},\ and\ \citenamefont
  {Cronin}}]{Parrilla-Gutierrez2020}%
  \BibitemOpen
  \bibfield  {author} {\bibinfo {author} {\bibfnamefont {J.~M.}\ \bibnamefont
  {Parrilla-Gutierrez}}, \bibinfo {author} {\bibfnamefont {A.}~\bibnamefont
  {Sharma}}, \bibinfo {author} {\bibfnamefont {S.}~\bibnamefont {Tsuda}},
  \bibinfo {author} {\bibfnamefont {G.~J.}\ \bibnamefont {Cooper}}, \bibinfo
  {author} {\bibfnamefont {G.}~\bibnamefont {Aragon-Camarasa}}, \bibinfo
  {author} {\bibfnamefont {K.}~\bibnamefont {Donkers}},\ and\ \bibinfo {author}
  {\bibfnamefont {L.}~\bibnamefont {Cronin}},\ }\href
  {https://doi.org/10.1038/s41467-020-15190-3} {\bibfield  {journal} {\bibinfo
  {journal} {Nature Communications}\ }\textbf {\bibinfo {volume} {11}},\
  \bibinfo {pages} {1} (\bibinfo {year} {2020})}\BibitemShut {NoStop}%
\bibitem [{\citenamefont {Wang}\ \emph {et~al.}(2016)\citenamefont {Wang},
  \citenamefont {Gold}, \citenamefont {Tompkins}, \citenamefont {Heymann},
  \citenamefont {Harrington},\ and\ \citenamefont {Fraden}}]{Wang2016}%
  \BibitemOpen
  \bibfield  {author} {\bibinfo {author} {\bibfnamefont {A.}~\bibnamefont
  {Wang}}, \bibinfo {author} {\bibfnamefont {J.}~\bibnamefont {Gold}}, \bibinfo
  {author} {\bibfnamefont {N.}~\bibnamefont {Tompkins}}, \bibinfo {author}
  {\bibfnamefont {M.}~\bibnamefont {Heymann}}, \bibinfo {author} {\bibfnamefont
  {K.}~\bibnamefont {Harrington}},\ and\ \bibinfo {author} {\bibfnamefont
  {S.}~\bibnamefont {Fraden}},\ }\href
  {https://doi.org/10.1140/epjst/e2016-02622-y} {\bibfield  {journal} {\bibinfo
   {journal} {The European Physical Journal Special Topics}\ }\textbf {\bibinfo
  {volume} {225}},\ \bibinfo {pages} {211} (\bibinfo {year}
  {2016})}\BibitemShut {NoStop}%
\bibitem [{\citenamefont {Yoshida}\ \emph {et~al.}(1996)\citenamefont
  {Yoshida}, \citenamefont {Takahashi}, \citenamefont {Yamaguchi},\ and\
  \citenamefont {Ichijo}}]{Yoshida1996}%
  \BibitemOpen
  \bibfield  {author} {\bibinfo {author} {\bibfnamefont {R.}~\bibnamefont
  {Yoshida}}, \bibinfo {author} {\bibfnamefont {T.}~\bibnamefont {Takahashi}},
  \bibinfo {author} {\bibfnamefont {T.}~\bibnamefont {Yamaguchi}},\ and\
  \bibinfo {author} {\bibfnamefont {H.}~\bibnamefont {Ichijo}},\ }\href
  {https://doi.org/10.1021/ja9602511} {\bibfield  {journal} {\bibinfo
  {journal} {Journal of the American Chemical Society}\ }\textbf {\bibinfo
  {volume} {118}},\ \bibinfo {pages} {5134} (\bibinfo {year}
  {1996})}\BibitemShut {NoStop}%
\bibitem [{\citenamefont {Homma}\ \emph {et~al.}(2017)\citenamefont {Homma},
  \citenamefont {Masuda}, \citenamefont {Akimoto}, \citenamefont {Nagase},
  \citenamefont {Itoga}, \citenamefont {Okano},\ and\ \citenamefont
  {Yoshida}}]{Homma2017}%
  \BibitemOpen
  \bibfield  {author} {\bibinfo {author} {\bibfnamefont {K.}~\bibnamefont
  {Homma}}, \bibinfo {author} {\bibfnamefont {T.}~\bibnamefont {Masuda}},
  \bibinfo {author} {\bibfnamefont {A.~M.}\ \bibnamefont {Akimoto}}, \bibinfo
  {author} {\bibfnamefont {K.}~\bibnamefont {Nagase}}, \bibinfo {author}
  {\bibfnamefont {K.}~\bibnamefont {Itoga}}, \bibinfo {author} {\bibfnamefont
  {T.}~\bibnamefont {Okano}},\ and\ \bibinfo {author} {\bibfnamefont
  {R.}~\bibnamefont {Yoshida}},\ }\href
  {https://doi.org/10.1002/smll.201700041} {\bibfield  {journal} {\bibinfo
  {journal} {Small}\ }\textbf {\bibinfo {volume} {13}},\ \bibinfo {pages} {1}
  (\bibinfo {year} {2017})}\BibitemShut {NoStop}%
\bibitem [{\citenamefont {Li}\ \emph {et~al.}(2019)\citenamefont {Li},
  \citenamefont {Li}, \citenamefont {Zheng},\ and\ \citenamefont
  {Ding}}]{Li2019}%
  \BibitemOpen
  \bibfield  {author} {\bibinfo {author} {\bibfnamefont {J.}~\bibnamefont
  {Li}}, \bibinfo {author} {\bibfnamefont {X.}~\bibnamefont {Li}}, \bibinfo
  {author} {\bibfnamefont {Z.}~\bibnamefont {Zheng}},\ and\ \bibinfo {author}
  {\bibfnamefont {X.}~\bibnamefont {Ding}},\ }\href
  {https://doi.org/10.1039/c9ra02340b} {\bibfield  {journal} {\bibinfo
  {journal} {RSC Advances}\ }\textbf {\bibinfo {volume} {9}},\ \bibinfo {pages}
  {13168} (\bibinfo {year} {2019})}\BibitemShut {NoStop}%
\bibitem [{\citenamefont {Mallphanov}\ and\ \citenamefont
  {Vanag}(2020)}]{Mallphanov2020}%
  \BibitemOpen
  \bibfield  {author} {\bibinfo {author} {\bibfnamefont {I.~L.}\ \bibnamefont
  {Mallphanov}}\ and\ \bibinfo {author} {\bibfnamefont {V.~K.}\ \bibnamefont
  {Vanag}},\ }\href {https://doi.org/10.1021/acs.jpca.9b09127} {\bibfield
  {journal} {\bibinfo  {journal} {Journal of Physical Chemistry A}\ }\textbf
  {\bibinfo {volume} {124}},\ \bibinfo {pages} {272} (\bibinfo {year}
  {2020})}\BibitemShut {NoStop}%
\bibitem [{\citenamefont {Ahmadi}\ and\ \citenamefont
  {Seiffert}(2020)}]{Ahmadi2020}%
  \BibitemOpen
  \bibfield  {author} {\bibinfo {author} {\bibfnamefont {M.}~\bibnamefont
  {Ahmadi}}\ and\ \bibinfo {author} {\bibfnamefont {S.}~\bibnamefont
  {Seiffert}},\ }\href {https://doi.org/10.1039/d0cp02429e} {\bibfield
  {journal} {\bibinfo  {journal} {Physical Chemistry Chemical Physics}\
  }\textbf {\bibinfo {volume} {22}},\ \bibinfo {pages} {14965} (\bibinfo {year}
  {2020})}\BibitemShut {NoStop}%
\bibitem [{\citenamefont {Cui}\ \emph {et~al.}(2020)\citenamefont {Cui},
  \citenamefont {Chen},\ and\ \citenamefont {Chen}}]{Cui2020}%
  \BibitemOpen
  \bibfield  {author} {\bibinfo {author} {\bibfnamefont {R.~F.}\ \bibnamefont
  {Cui}}, \bibinfo {author} {\bibfnamefont {Q.~H.}\ \bibnamefont {Chen}},\ and\
  \bibinfo {author} {\bibfnamefont {J.~X.}\ \bibnamefont {Chen}},\ }\href
  {https://doi.org/10.1039/d0nr01211d} {\bibfield  {journal} {\bibinfo
  {journal} {Nanoscale}\ }\textbf {\bibinfo {volume} {12}},\ \bibinfo {pages}
  {12275} (\bibinfo {year} {2020})}\BibitemShut {NoStop}%
\bibitem [{\citenamefont {Litschel}\ \emph {et~al.}(2018)\citenamefont
  {Litschel}, \citenamefont {Norton}, \citenamefont {Tserunyan},\ and\
  \citenamefont {Fraden}}]{Litschel2018}%
  \BibitemOpen
  \bibfield  {author} {\bibinfo {author} {\bibfnamefont {T.}~\bibnamefont
  {Litschel}}, \bibinfo {author} {\bibfnamefont {M.~M.}\ \bibnamefont
  {Norton}}, \bibinfo {author} {\bibfnamefont {V.}~\bibnamefont {Tserunyan}},\
  and\ \bibinfo {author} {\bibfnamefont {S.}~\bibnamefont {Fraden}},\ }\href
  {https://doi.org/10.1039/c7lc01187c} {\bibfield  {journal} {\bibinfo
  {journal} {Lab on a Chip}\ }\textbf {\bibinfo {volume} {18}},\ \bibinfo
  {pages} {714} (\bibinfo {year} {2018})}\BibitemShut {NoStop}%
\bibitem [{\citenamefont {Ren}\ \emph {et~al.}(2017)\citenamefont {Ren},
  \citenamefont {Wang}, \citenamefont {Pan}, \citenamefont {Gao}, \citenamefont
  {Liu},\ and\ \citenamefont {Epstein}}]{Ren2017}%
  \BibitemOpen
  \bibfield  {author} {\bibinfo {author} {\bibfnamefont {L.}~\bibnamefont
  {Ren}}, \bibinfo {author} {\bibfnamefont {M.}~\bibnamefont {Wang}}, \bibinfo
  {author} {\bibfnamefont {C.}~\bibnamefont {Pan}}, \bibinfo {author}
  {\bibfnamefont {Q.}~\bibnamefont {Gao}}, \bibinfo {author} {\bibfnamefont
  {Y.}~\bibnamefont {Liu}},\ and\ \bibinfo {author} {\bibfnamefont {I.~R.}\
  \bibnamefont {Epstein}},\ }\href {https://doi.org/10.1073/pnas.1704094114}
  {\bibfield  {journal} {\bibinfo  {journal} {Proceedings of the National
  Academy of Sciences}\ }\textbf {\bibinfo {volume} {114}},\ \bibinfo {pages}
  {8704} (\bibinfo {year} {2017})}\BibitemShut {NoStop}%
\bibitem [{\citenamefont {Ren}\ \emph {et~al.}(2020)\citenamefont {Ren},
  \citenamefont {Yuan}, \citenamefont {Gao}, \citenamefont {Teng},
  \citenamefont {Wang},\ and\ \citenamefont {Epstein}}]{Ren2020}%
  \BibitemOpen
  \bibfield  {author} {\bibinfo {author} {\bibfnamefont {L.}~\bibnamefont
  {Ren}}, \bibinfo {author} {\bibfnamefont {L.}~\bibnamefont {Yuan}}, \bibinfo
  {author} {\bibfnamefont {Q.}~\bibnamefont {Gao}}, \bibinfo {author}
  {\bibfnamefont {R.}~\bibnamefont {Teng}}, \bibinfo {author} {\bibfnamefont
  {J.}~\bibnamefont {Wang}},\ and\ \bibinfo {author} {\bibfnamefont {I.~R.}\
  \bibnamefont {Epstein}},\ }\href {https://doi.org/10.1126/sciadv.aaz9125}
  {\bibfield  {journal} {\bibinfo  {journal} {Science Advances}\ }\textbf
  {\bibinfo {volume} {6}},\ \bibinfo {pages} {1} (\bibinfo {year}
  {2020})}\BibitemShut {NoStop}%
\bibitem [{\citenamefont {Nishide}\ and\ \citenamefont
  {Ishihara}(2022)}]{Nishide2022}%
  \BibitemOpen
  \bibfield  {author} {\bibinfo {author} {\bibfnamefont {R.}~\bibnamefont
  {Nishide}}\ and\ \bibinfo {author} {\bibfnamefont {S.}~\bibnamefont
  {Ishihara}},\ }\href {https://doi.org/10.1103/PhysRevLett.128.224101}
  {\bibfield  {journal} {\bibinfo  {journal} {Physical Review Letters}\
  }\textbf {\bibinfo {volume} {128}},\ \bibinfo {pages} {224101} (\bibinfo
  {year} {2022})}\BibitemShut {NoStop}%
\bibitem [{\citenamefont {Frank}\ \emph {et~al.}(2019)\citenamefont {Frank},
  \citenamefont {Guven}, \citenamefont {Kardar},\ and\ \citenamefont
  {Shackleton}}]{Frank2019}%
  \BibitemOpen
  \bibfield  {author} {\bibinfo {author} {\bibfnamefont {J.~R.}\ \bibnamefont
  {Frank}}, \bibinfo {author} {\bibfnamefont {J.}~\bibnamefont {Guven}},
  \bibinfo {author} {\bibfnamefont {M.}~\bibnamefont {Kardar}},\ and\ \bibinfo
  {author} {\bibfnamefont {H.}~\bibnamefont {Shackleton}},\ }\href
  {https://doi.org/10.1209/0295-5075/127/48001} {\bibfield  {journal} {\bibinfo
   {journal} {EPL (Europhysics Letters)}\ }\textbf {\bibinfo {volume} {127}},\
  \bibinfo {pages} {48001} (\bibinfo {year} {2019})}\BibitemShut {NoStop}%
\bibitem [{\citenamefont {Krause}\ \emph {et~al.}(2021)\citenamefont {Krause},
  \citenamefont {Gaffney}, \citenamefont {Maini},\ and\ \citenamefont
  {Klika}}]{Krause2021}%
  \BibitemOpen
  \bibfield  {author} {\bibinfo {author} {\bibfnamefont {A.~L.}\ \bibnamefont
  {Krause}}, \bibinfo {author} {\bibfnamefont {E.~A.}\ \bibnamefont {Gaffney}},
  \bibinfo {author} {\bibfnamefont {P.~K.}\ \bibnamefont {Maini}},\ and\
  \bibinfo {author} {\bibfnamefont {V.}~\bibnamefont {Klika}},\ }\bibfield
  {journal} {\bibinfo  {journal} {Philosophical Transactions of the Royal
  Society A: Mathematical, Physical and Engineering Sciences}\ }\textbf
  {\bibinfo {volume} {379}},\ \href {https://doi.org/10.1098/rsta.2020.0268}
  {10.1098/rsta.2020.0268} (\bibinfo {year} {2021})\BibitemShut {NoStop}%
\bibitem [{\citenamefont {Coyte}\ \emph {et~al.}(2017)\citenamefont {Coyte},
  \citenamefont {Tabuteau}, \citenamefont {Gaffney}, \citenamefont {Foster},\
  and\ \citenamefont {Durham}}]{Coyte2017}%
  \BibitemOpen
  \bibfield  {author} {\bibinfo {author} {\bibfnamefont {K.~Z.}\ \bibnamefont
  {Coyte}}, \bibinfo {author} {\bibfnamefont {H.}~\bibnamefont {Tabuteau}},
  \bibinfo {author} {\bibfnamefont {E.~A.}\ \bibnamefont {Gaffney}}, \bibinfo
  {author} {\bibfnamefont {K.~R.}\ \bibnamefont {Foster}},\ and\ \bibinfo
  {author} {\bibfnamefont {W.~M.}\ \bibnamefont {Durham}},\ }\href
  {https://doi.org/10.1073/pnas.1525228113} {\bibfield  {journal} {\bibinfo
  {journal} {Proceedings of the National Academy of Sciences}\ }\textbf
  {\bibinfo {volume} {114}},\ \bibinfo {pages} {E161} (\bibinfo {year}
  {2017})}\BibitemShut {NoStop}%
\bibitem [{\citenamefont {de~Anna}\ \emph {et~al.}(2021)\citenamefont
  {de~Anna}, \citenamefont {Pahlavan}, \citenamefont {Yawata}, \citenamefont
  {Stocker},\ and\ \citenamefont {Juanes}}]{DeAnna2021}%
  \BibitemOpen
  \bibfield  {author} {\bibinfo {author} {\bibfnamefont {P.}~\bibnamefont
  {de~Anna}}, \bibinfo {author} {\bibfnamefont {A.~A.}\ \bibnamefont
  {Pahlavan}}, \bibinfo {author} {\bibfnamefont {Y.}~\bibnamefont {Yawata}},
  \bibinfo {author} {\bibfnamefont {R.}~\bibnamefont {Stocker}},\ and\ \bibinfo
  {author} {\bibfnamefont {R.}~\bibnamefont {Juanes}},\ }\href
  {https://doi.org/10.1038/s41567-020-1002-x} {\bibfield  {journal} {\bibinfo
  {journal} {Nature Physics}\ }\textbf {\bibinfo {volume} {17}},\ \bibinfo
  {pages} {68} (\bibinfo {year} {2021})}\BibitemShut {NoStop}%
\bibitem [{\citenamefont {Dehkharghani}\ \emph {et~al.}(2019)\citenamefont
  {Dehkharghani}, \citenamefont {Waisbord}, \citenamefont {Dunkel},\ and\
  \citenamefont {Guasto}}]{Dehkharghani2019}%
  \BibitemOpen
  \bibfield  {author} {\bibinfo {author} {\bibfnamefont {A.}~\bibnamefont
  {Dehkharghani}}, \bibinfo {author} {\bibfnamefont {N.}~\bibnamefont
  {Waisbord}}, \bibinfo {author} {\bibfnamefont {J.}~\bibnamefont {Dunkel}},\
  and\ \bibinfo {author} {\bibfnamefont {J.~S.}\ \bibnamefont {Guasto}},\
  }\href {https://doi.org/10.1073/pnas.1819613116} {\bibfield  {journal}
  {\bibinfo  {journal} {Proceedings of the National Academy of Sciences}\
  }\textbf {\bibinfo {volume} {116}},\ \bibinfo {pages} {11119} (\bibinfo
  {year} {2019})}\BibitemShut {NoStop}%
\bibitem [{\citenamefont {Atis}\ \emph {et~al.}(2013)\citenamefont {Atis},
  \citenamefont {Saha}, \citenamefont {Auradou}, \citenamefont {Salin},\ and\
  \citenamefont {Talon}}]{Atis2013}%
  \BibitemOpen
  \bibfield  {author} {\bibinfo {author} {\bibfnamefont {S.}~\bibnamefont
  {Atis}}, \bibinfo {author} {\bibfnamefont {S.}~\bibnamefont {Saha}}, \bibinfo
  {author} {\bibfnamefont {H.}~\bibnamefont {Auradou}}, \bibinfo {author}
  {\bibfnamefont {D.}~\bibnamefont {Salin}},\ and\ \bibinfo {author}
  {\bibfnamefont {L.}~\bibnamefont {Talon}},\ }\href
  {https://doi.org/10.1103/PhysRevLett.110.148301} {\bibfield  {journal}
  {\bibinfo  {journal} {Physical Review Letters}\ }\textbf {\bibinfo {volume}
  {110}},\ \bibinfo {pages} {1} (\bibinfo {year} {2013})}\BibitemShut {NoStop}%
\bibitem [{\citenamefont {Goldbeter}(2017)}]{Goldbeter2017}%
  \BibitemOpen
  \bibfield  {author} {\bibinfo {author} {\bibfnamefont {A.}~\bibnamefont
  {Goldbeter}},\ }\bibfield  {journal} {\bibinfo  {journal} {Chaos}\ }\textbf
  {\bibinfo {volume} {27}},\ \href {https://doi.org/10.1063/1.4990783}
  {10.1063/1.4990783} (\bibinfo {year} {2017})\BibitemShut {NoStop}%
\bibitem [{\citenamefont {Goldbeter}(2018)}]{Goldbeter2018}%
  \BibitemOpen
  \bibfield  {author} {\bibinfo {author} {\bibfnamefont {A.}~\bibnamefont
  {Goldbeter}},\ }\href {https://doi.org/10.1098/rsta.2017.0376} {\bibfield
  {journal} {\bibinfo  {journal} {Philosophical Transactions of the Royal
  Society A: Mathematical, Physical and Engineering Sciences}\ }\textbf
  {\bibinfo {volume} {376}},\ \bibinfo {pages} {20170376} (\bibinfo {year}
  {2018})}\BibitemShut {NoStop}%
\bibitem [{\citenamefont {de~Anna}\ \emph {et~al.}(2014)\citenamefont
  {de~Anna}, \citenamefont {Dentz}, \citenamefont {Tartakovsky},\ and\
  \citenamefont {Le~Borgne}}]{DeAnna2014a}%
  \BibitemOpen
  \bibfield  {author} {\bibinfo {author} {\bibfnamefont {P.}~\bibnamefont
  {de~Anna}}, \bibinfo {author} {\bibfnamefont {M.}~\bibnamefont {Dentz}},
  \bibinfo {author} {\bibfnamefont {A.}~\bibnamefont {Tartakovsky}},\ and\
  \bibinfo {author} {\bibfnamefont {T.}~\bibnamefont {Le~Borgne}},\ }\href
  {https://doi.org/10.1002/2014GL060068} {\bibfield  {journal} {\bibinfo
  {journal} {Geophysical Research Letters}\ }\textbf {\bibinfo {volume} {41}},\
  \bibinfo {pages} {4586} (\bibinfo {year} {2014})}\BibitemShut {NoStop}%
\bibitem [{\citenamefont {Valocchi}\ \emph {et~al.}(2019)\citenamefont
  {Valocchi}, \citenamefont {Bolster},\ and\ \citenamefont
  {Werth}}]{Valocchi2019b}%
  \BibitemOpen
  \bibfield  {author} {\bibinfo {author} {\bibfnamefont {A.~J.}\ \bibnamefont
  {Valocchi}}, \bibinfo {author} {\bibfnamefont {D.}~\bibnamefont {Bolster}},\
  and\ \bibinfo {author} {\bibfnamefont {C.~J.}\ \bibnamefont {Werth}},\ }\href
  {https://doi.org/10.1007/s11242-018-1204-1} {\bibfield  {journal} {\bibinfo
  {journal} {Transport in Porous Media}\ }\textbf {\bibinfo {volume} {130}},\
  \bibinfo {pages} {157} (\bibinfo {year} {2019})}\BibitemShut {NoStop}%
\bibitem [{\citenamefont {Yoon}\ \emph {et~al.}(2021)\citenamefont {Yoon},
  \citenamefont {Dentz},\ and\ \citenamefont {Kang}}]{Yoon2021}%
  \BibitemOpen
  \bibfield  {author} {\bibinfo {author} {\bibfnamefont {S.}~\bibnamefont
  {Yoon}}, \bibinfo {author} {\bibfnamefont {M.}~\bibnamefont {Dentz}},\ and\
  \bibinfo {author} {\bibfnamefont {P.~K.}\ \bibnamefont {Kang}},\ }\href
  {https://doi.org/10.1017/jfm.2021.208} {\bibfield  {journal} {\bibinfo
  {journal} {Journal of Fluid Mechanics}\ }\textbf {\bibinfo {volume} {916}},\
  \bibinfo {pages} {1} (\bibinfo {year} {2021})}\BibitemShut {NoStop}%
\bibitem [{\citenamefont {Izumoto}\ \emph {et~al.}(2022)\citenamefont
  {Izumoto}, \citenamefont {Huisman}, \citenamefont {Zimmermann}, \citenamefont
  {Heyman}, \citenamefont {Gomez}, \citenamefont {Tabuteau}, \citenamefont
  {Laniel}, \citenamefont {Vereecken}, \citenamefont {M{\'{e}}heust},\ and\
  \citenamefont {Le~Borgne}}]{Izumoto2022}%
  \BibitemOpen
  \bibfield  {author} {\bibinfo {author} {\bibfnamefont {S.}~\bibnamefont
  {Izumoto}}, \bibinfo {author} {\bibfnamefont {J.~A.}\ \bibnamefont
  {Huisman}}, \bibinfo {author} {\bibfnamefont {E.}~\bibnamefont {Zimmermann}},
  \bibinfo {author} {\bibfnamefont {J.}~\bibnamefont {Heyman}}, \bibinfo
  {author} {\bibfnamefont {F.}~\bibnamefont {Gomez}}, \bibinfo {author}
  {\bibfnamefont {H.}~\bibnamefont {Tabuteau}}, \bibinfo {author}
  {\bibfnamefont {R.}~\bibnamefont {Laniel}}, \bibinfo {author} {\bibfnamefont
  {H.}~\bibnamefont {Vereecken}}, \bibinfo {author} {\bibfnamefont
  {Y.}~\bibnamefont {M{\'{e}}heust}},\ and\ \bibinfo {author} {\bibfnamefont
  {T.}~\bibnamefont {Le~Borgne}},\ }\href
  {https://doi.org/10.1021/acs.est.1c07742} {\bibfield  {journal} {\bibinfo
  {journal} {Environmental Science {\&} Technology}\ }\textbf {\bibinfo
  {volume} {56}},\ \bibinfo {pages} {4998} (\bibinfo {year}
  {2022})}\BibitemShut {NoStop}%
\bibitem [{\citenamefont {Soulaine}\ \emph {et~al.}(2018)\citenamefont
  {Soulaine}, \citenamefont {Roman}, \citenamefont {Kovscek},\ and\
  \citenamefont {Tchelepi}}]{Soulaine2018}%
  \BibitemOpen
  \bibfield  {author} {\bibinfo {author} {\bibfnamefont {C.}~\bibnamefont
  {Soulaine}}, \bibinfo {author} {\bibfnamefont {S.}~\bibnamefont {Roman}},
  \bibinfo {author} {\bibfnamefont {A.}~\bibnamefont {Kovscek}},\ and\ \bibinfo
  {author} {\bibfnamefont {H.~A.}\ \bibnamefont {Tchelepi}},\ }\href
  {https://doi.org/10.1017/jfm.2018.655} {\bibfield  {journal} {\bibinfo
  {journal} {Journal of Fluid Mechanics}\ }\textbf {\bibinfo {volume} {855}},\
  \bibinfo {pages} {616} (\bibinfo {year} {2018})}\BibitemShut {NoStop}%
\bibitem [{\citenamefont {Sepulchre}\ and\ \citenamefont
  {Babloyantz}(1993)}]{Sepulchre1993}%
  \BibitemOpen
  \bibfield  {author} {\bibinfo {author} {\bibfnamefont {J.~A.}\ \bibnamefont
  {Sepulchre}}\ and\ \bibinfo {author} {\bibfnamefont {A.}~\bibnamefont
  {Babloyantz}},\ }\href {https://doi.org/10.1103/PhysRevE.48.187} {\bibfield
  {journal} {\bibinfo  {journal} {Physical Review E}\ }\textbf {\bibinfo
  {volume} {48}},\ \bibinfo {pages} {187} (\bibinfo {year} {1993})}\BibitemShut
  {NoStop}%
\bibitem [{\citenamefont {Marlow}\ \emph {et~al.}(1997)\citenamefont {Marlow},
  \citenamefont {Sasaki},\ and\ \citenamefont {Vasquez}}]{Marlow1997}%
  \BibitemOpen
  \bibfield  {author} {\bibinfo {author} {\bibfnamefont {M.}~\bibnamefont
  {Marlow}}, \bibinfo {author} {\bibfnamefont {Y.}~\bibnamefont {Sasaki}},\
  and\ \bibinfo {author} {\bibfnamefont {D.~A.}\ \bibnamefont {Vasquez}},\
  }\href {https://doi.org/10.1063/1.474883} {\bibfield  {journal} {\bibinfo
  {journal} {The Journal of Chemical Physics}\ }\textbf {\bibinfo {volume}
  {107}},\ \bibinfo {pages} {5205} (\bibinfo {year} {1997})}\BibitemShut
  {NoStop}%
\bibitem [{\citenamefont {Ginn}\ and\ \citenamefont
  {Steinbock}(2005)}]{Ginn2005}%
  \BibitemOpen
  \bibfield  {author} {\bibinfo {author} {\bibfnamefont {B.~T.}\ \bibnamefont
  {Ginn}}\ and\ \bibinfo {author} {\bibfnamefont {O.}~\bibnamefont
  {Steinbock}},\ }\href {https://doi.org/10.1103/PhysRevE.72.046109} {\bibfield
   {journal} {\bibinfo  {journal} {Physical Review E}\ }\textbf {\bibinfo
  {volume} {72}},\ \bibinfo {pages} {046109} (\bibinfo {year}
  {2005})}\BibitemShut {NoStop}%
\bibitem [{\citenamefont {Atis}\ \emph {et~al.}(2012)\citenamefont {Atis},
  \citenamefont {Saha}, \citenamefont {Auradou}, \citenamefont {Martin},
  \citenamefont {Rakotomalala}, \citenamefont {Talon},\ and\ \citenamefont
  {Salin}}]{Atis2012}%
  \BibitemOpen
  \bibfield  {author} {\bibinfo {author} {\bibfnamefont {S.}~\bibnamefont
  {Atis}}, \bibinfo {author} {\bibfnamefont {S.}~\bibnamefont {Saha}}, \bibinfo
  {author} {\bibfnamefont {H.}~\bibnamefont {Auradou}}, \bibinfo {author}
  {\bibfnamefont {J.}~\bibnamefont {Martin}}, \bibinfo {author} {\bibfnamefont
  {N.}~\bibnamefont {Rakotomalala}}, \bibinfo {author} {\bibfnamefont
  {L.}~\bibnamefont {Talon}},\ and\ \bibinfo {author} {\bibfnamefont
  {D.}~\bibnamefont {Salin}},\ }\href {https://doi.org/10.1063/1.4734489}
  {\bibfield  {journal} {\bibinfo  {journal} {Chaos: An Interdisciplinary
  Journal of Nonlinear Science}\ }\textbf {\bibinfo {volume} {22}},\ \bibinfo
  {pages} {037108} (\bibinfo {year} {2012})}\BibitemShut {NoStop}%
\bibitem [{\citenamefont {Chevalier}\ \emph {et~al.}(2017)\citenamefont
  {Chevalier}, \citenamefont {Dubey}, \citenamefont {Atis}, \citenamefont
  {Rosso}, \citenamefont {Salin},\ and\ \citenamefont {Talon}}]{Chevalier2017}%
  \BibitemOpen
  \bibfield  {author} {\bibinfo {author} {\bibfnamefont {T.}~\bibnamefont
  {Chevalier}}, \bibinfo {author} {\bibfnamefont {A.~K.}\ \bibnamefont
  {Dubey}}, \bibinfo {author} {\bibfnamefont {S.}~\bibnamefont {Atis}},
  \bibinfo {author} {\bibfnamefont {A.}~\bibnamefont {Rosso}}, \bibinfo
  {author} {\bibfnamefont {D.}~\bibnamefont {Salin}},\ and\ \bibinfo {author}
  {\bibfnamefont {L.}~\bibnamefont {Talon}},\ }\href
  {https://doi.org/10.1103/PhysRevE.95.042210} {\bibfield  {journal} {\bibinfo
  {journal} {Physical Review E}\ }\textbf {\bibinfo {volume} {95}},\ \bibinfo
  {pages} {1} (\bibinfo {year} {2017})}\BibitemShut {NoStop}%
\bibitem [{\citenamefont {Rolle}\ and\ \citenamefont
  {Le~Borgne}(2019)}]{Rolle2019}%
  \BibitemOpen
  \bibfield  {author} {\bibinfo {author} {\bibfnamefont {M.}~\bibnamefont
  {Rolle}}\ and\ \bibinfo {author} {\bibfnamefont {T.}~\bibnamefont
  {Le~Borgne}},\ }\href {https://doi.org/10.2138/rmg.2018.85.5} {\bibfield
  {journal} {\bibinfo  {journal} {Reviews in Mineralogy and Geochemistry}\
  }\textbf {\bibinfo {volume} {85}},\ \bibinfo {pages} {111} (\bibinfo {year}
  {2019})}\BibitemShut {NoStop}%
\bibitem [{\citenamefont {Dentz}\ \emph {et~al.}(2011)\citenamefont {Dentz},
  \citenamefont {Le~Borgne}, \citenamefont {Englert},\ and\ \citenamefont
  {Bijeljic}}]{Dentz2011}%
  \BibitemOpen
  \bibfield  {author} {\bibinfo {author} {\bibfnamefont {M.}~\bibnamefont
  {Dentz}}, \bibinfo {author} {\bibfnamefont {T.}~\bibnamefont {Le~Borgne}},
  \bibinfo {author} {\bibfnamefont {A.}~\bibnamefont {Englert}},\ and\ \bibinfo
  {author} {\bibfnamefont {B.}~\bibnamefont {Bijeljic}},\ }\href
  {https://doi.org/10.1016/j.jconhyd.2010.05.002} {\bibfield  {journal}
  {\bibinfo  {journal} {Journal of Contaminant Hydrology}\ }\textbf {\bibinfo
  {volume} {120-121}},\ \bibinfo {pages} {1} (\bibinfo {year}
  {2011})}\BibitemShut {NoStop}%
\bibitem [{\citenamefont {Haller}(2015)}]{Haller2015LagrangianStructures}%
  \BibitemOpen
  \bibfield  {author} {\bibinfo {author} {\bibfnamefont {G.}~\bibnamefont
  {Haller}},\ }\href {https://doi.org/10.1146/annurev-fluid-010313-141322}
  {\bibfield  {journal} {\bibinfo  {journal} {Annual Review of Fluid
  Mechanics}\ }\textbf {\bibinfo {volume} {47}},\ \bibinfo {pages} {137}
  (\bibinfo {year} {2015})}\BibitemShut {NoStop}%
\bibitem [{\citenamefont {Mahoney}\ \emph {et~al.}(2012)\citenamefont
  {Mahoney}, \citenamefont {Bargteil}, \citenamefont {Kingsbury}, \citenamefont
  {Mitchell},\ and\ \citenamefont {Solomon}}]{Mahoney2012}%
  \BibitemOpen
  \bibfield  {author} {\bibinfo {author} {\bibfnamefont {J.}~\bibnamefont
  {Mahoney}}, \bibinfo {author} {\bibfnamefont {D.}~\bibnamefont {Bargteil}},
  \bibinfo {author} {\bibfnamefont {M.}~\bibnamefont {Kingsbury}}, \bibinfo
  {author} {\bibfnamefont {K.}~\bibnamefont {Mitchell}},\ and\ \bibinfo
  {author} {\bibfnamefont {T.}~\bibnamefont {Solomon}},\ }\href
  {https://doi.org/10.1209/0295-5075/98/44005} {\bibfield  {journal} {\bibinfo
  {journal} {EPL (Europhysics Letters)}\ }\textbf {\bibinfo {volume} {98}},\
  \bibinfo {pages} {44005} (\bibinfo {year} {2012})}\BibitemShut {NoStop}%
\bibitem [{\citenamefont {Mitchell}\ and\ \citenamefont
  {Mahoney}(2012)}]{Mitchell2012}%
  \BibitemOpen
  \bibfield  {author} {\bibinfo {author} {\bibfnamefont {K.~A.}\ \bibnamefont
  {Mitchell}}\ and\ \bibinfo {author} {\bibfnamefont {J.~R.}\ \bibnamefont
  {Mahoney}},\ }\bibfield  {journal} {\bibinfo  {journal} {Chaos}\ }\textbf
  {\bibinfo {volume} {22}},\ \href {https://doi.org/10.1063/1.4746039}
  {10.1063/1.4746039} (\bibinfo {year} {2012})\BibitemShut {NoStop}%
\bibitem [{\citenamefont {Berman}\ \emph {et~al.}(2021)\citenamefont {Berman},
  \citenamefont {Buggeln}, \citenamefont {Brantley}, \citenamefont {Mitchell},\
  and\ \citenamefont {Solomon}}]{Berman2021TransportFlows}%
  \BibitemOpen
  \bibfield  {author} {\bibinfo {author} {\bibfnamefont {S.~A.}\ \bibnamefont
  {Berman}}, \bibinfo {author} {\bibfnamefont {J.}~\bibnamefont {Buggeln}},
  \bibinfo {author} {\bibfnamefont {D.~A.}\ \bibnamefont {Brantley}}, \bibinfo
  {author} {\bibfnamefont {K.~A.}\ \bibnamefont {Mitchell}},\ and\ \bibinfo
  {author} {\bibfnamefont {T.~H.}\ \bibnamefont {Solomon}},\ }\bibfield
  {journal} {\bibinfo  {journal} {Physical Review Fluids}\ }\textbf {\bibinfo
  {volume} {6}},\ \href {https://doi.org/10.1103/PhysRevFluids.6.L012501}
  {10.1103/PhysRevFluids.6.L012501} (\bibinfo {year} {2021})\BibitemShut
  {NoStop}%
\bibitem [{\citenamefont {Izumoto}\ \emph {et~al.}(2023)\citenamefont
  {Izumoto}, \citenamefont {Heyman}, \citenamefont {Huisman}, \citenamefont
  {De~Vriendt}, \citenamefont {Soulaine}, \citenamefont {Gomez}, \citenamefont
  {Tabuteau}, \citenamefont {M{\'{e}}heust},\ and\ \citenamefont
  {Le~Borgne}}]{izumoto2023b}%
  \BibitemOpen
  \bibfield  {author} {\bibinfo {author} {\bibfnamefont {S.}~\bibnamefont
  {Izumoto}}, \bibinfo {author} {\bibfnamefont {J.}~\bibnamefont {Heyman}},
  \bibinfo {author} {\bibfnamefont {J.~A.}\ \bibnamefont {Huisman}}, \bibinfo
  {author} {\bibfnamefont {K.}~\bibnamefont {De~Vriendt}}, \bibinfo {author}
  {\bibfnamefont {C.}~\bibnamefont {Soulaine}}, \bibinfo {author}
  {\bibfnamefont {F.}~\bibnamefont {Gomez}}, \bibinfo {author} {\bibfnamefont
  {H.}~\bibnamefont {Tabuteau}}, \bibinfo {author} {\bibfnamefont
  {Y.}~\bibnamefont {M{\'{e}}heust}},\ and\ \bibinfo {author} {\bibfnamefont
  {T.}~\bibnamefont {Le~Borgne}},\ }\bibfield  {journal} {\bibinfo  {journal}
  {Water Resources Research}\ }\textbf {\bibinfo {volume} {59}},\ \href
  {https://doi.org/10.1029/2023WR034749} {10.1029/2023WR034749} (\bibinfo
  {year} {2023})\BibitemShut {NoStop}%
\bibitem [{\citenamefont {Metcalfe}\ \emph {et~al.}(2022)\citenamefont
  {Metcalfe}, \citenamefont {Lester},\ and\ \citenamefont
  {Trefry}}]{Metcalfe2022}%
  \BibitemOpen
  \bibfield  {author} {\bibinfo {author} {\bibfnamefont {G.}~\bibnamefont
  {Metcalfe}}, \bibinfo {author} {\bibfnamefont {D.}~\bibnamefont {Lester}},\
  and\ \bibinfo {author} {\bibfnamefont {M.}~\bibnamefont {Trefry}},\
  }\bibfield  {journal} {\bibinfo  {journal} {Transport in Porous Media}\
  }\href {https://doi.org/10.1007/s11242-022-01811-6}
  {10.1007/s11242-022-01811-6} (\bibinfo {year} {2022})\BibitemShut {NoStop}%
\bibitem [{\citenamefont {Yortsos}\ and\ \citenamefont
  {Zeybek}(1988)}]{Yortsos1988}%
  \BibitemOpen
  \bibfield  {author} {\bibinfo {author} {\bibfnamefont {Y.~C.}\ \bibnamefont
  {Yortsos}}\ and\ \bibinfo {author} {\bibfnamefont {M.}~\bibnamefont
  {Zeybek}},\ }\href {https://doi.org/10.1063/1.866918} {\bibfield  {journal}
  {\bibinfo  {journal} {Physics of Fluids}\ }\textbf {\bibinfo {volume} {31}},\
  \bibinfo {pages} {3511} (\bibinfo {year} {1988})}\BibitemShut {NoStop}%
\bibitem [{\citenamefont {Ghesmat}\ and\ \citenamefont
  {Azaiez}(2008)}]{Ghesmat2008}%
  \BibitemOpen
  \bibfield  {author} {\bibinfo {author} {\bibfnamefont {K.}~\bibnamefont
  {Ghesmat}}\ and\ \bibinfo {author} {\bibfnamefont {J.}~\bibnamefont
  {Azaiez}},\ }\href {https://doi.org/10.1007/s11242-007-9171-y} {\bibfield
  {journal} {\bibinfo  {journal} {Transport in Porous Media}\ }\textbf
  {\bibinfo {volume} {73}},\ \bibinfo {pages} {297} (\bibinfo {year}
  {2008})}\BibitemShut {NoStop}%
\bibitem [{\citenamefont {Budroni}\ \emph {et~al.}(2016)\citenamefont
  {Budroni}, \citenamefont {Lemaigre}, \citenamefont {Escala}, \citenamefont
  {Mu{\~{n}}uzuri},\ and\ \citenamefont {De~Wit}}]{Budroni2016a}%
  \BibitemOpen
  \bibfield  {author} {\bibinfo {author} {\bibfnamefont {M.~A.}\ \bibnamefont
  {Budroni}}, \bibinfo {author} {\bibfnamefont {L.}~\bibnamefont {Lemaigre}},
  \bibinfo {author} {\bibfnamefont {D.~M.}\ \bibnamefont {Escala}}, \bibinfo
  {author} {\bibfnamefont {A.~P.}\ \bibnamefont {Mu{\~{n}}uzuri}},\ and\
  \bibinfo {author} {\bibfnamefont {A.}~\bibnamefont {De~Wit}},\ }\href
  {https://doi.org/10.1021/acs.jpca.5b10802} {\bibfield  {journal} {\bibinfo
  {journal} {Journal of Physical Chemistry A}\ }\textbf {\bibinfo {volume}
  {120}},\ \bibinfo {pages} {851} (\bibinfo {year} {2016})}\BibitemShut
  {NoStop}%
\bibitem [{\citenamefont {Pe{\~{n}}a}\ and\ \citenamefont
  {P{\'{e}}rez-Garc{\'{i}}a}(2001)}]{Pena2001}%
  \BibitemOpen
  \bibfield  {author} {\bibinfo {author} {\bibfnamefont {B.}~\bibnamefont
  {Pe{\~{n}}a}}\ and\ \bibinfo {author} {\bibfnamefont {C.}~\bibnamefont
  {P{\'{e}}rez-Garc{\'{i}}a}},\ }\href
  {https://doi.org/10.1103/PhysRevE.64.056213} {\bibfield  {journal} {\bibinfo
  {journal} {Physical Review E}\ }\textbf {\bibinfo {volume} {64}},\ \bibinfo
  {pages} {056213} (\bibinfo {year} {2001})}\BibitemShut {NoStop}%
\bibitem [{\citenamefont {Andres{\'{e}}n}\ \emph {et~al.}(1999)\citenamefont
  {Andres{\'{e}}n}, \citenamefont {Bache}, \citenamefont {Mosekilde},
  \citenamefont {Dewel},\ and\ \citenamefont {Borckmanns}}]{Andresen1999}%
  \BibitemOpen
  \bibfield  {author} {\bibinfo {author} {\bibfnamefont {P.}~\bibnamefont
  {Andres{\'{e}}n}}, \bibinfo {author} {\bibfnamefont {M.}~\bibnamefont
  {Bache}}, \bibinfo {author} {\bibfnamefont {E.}~\bibnamefont {Mosekilde}},
  \bibinfo {author} {\bibfnamefont {G.}~\bibnamefont {Dewel}},\ and\ \bibinfo
  {author} {\bibfnamefont {P.}~\bibnamefont {Borckmanns}},\ }\href
  {https://doi.org/10.1103/PhysRevE.60.297} {\bibfield  {journal} {\bibinfo
  {journal} {Physical Review E}\ }\textbf {\bibinfo {volume} {60}},\ \bibinfo
  {pages} {297} (\bibinfo {year} {1999})}\BibitemShut {NoStop}%
\bibitem [{\citenamefont {Bamforth}\ \emph {et~al.}(2000)\citenamefont
  {Bamforth}, \citenamefont {Kalliadasis}, \citenamefont {Merkin},\ and\
  \citenamefont {Scott}}]{Bamforth2000}%
  \BibitemOpen
  \bibfield  {author} {\bibinfo {author} {\bibfnamefont {J.~R.}\ \bibnamefont
  {Bamforth}}, \bibinfo {author} {\bibfnamefont {S.}~\bibnamefont
  {Kalliadasis}}, \bibinfo {author} {\bibfnamefont {J.~H.}\ \bibnamefont
  {Merkin}},\ and\ \bibinfo {author} {\bibfnamefont {S.~K.}\ \bibnamefont
  {Scott}},\ }\href {https://doi.org/10.1039/b004552g} {\bibfield  {journal}
  {\bibinfo  {journal} {Physical Chemistry Chemical Physics}\ }\textbf
  {\bibinfo {volume} {2}},\ \bibinfo {pages} {4013} (\bibinfo {year}
  {2000})}\BibitemShut {NoStop}%
\bibitem [{\citenamefont {K{\ae}rn}\ and\ \citenamefont
  {Menzinger}(2000)}]{Kaern2000}%
  \BibitemOpen
  \bibfield  {author} {\bibinfo {author} {\bibfnamefont {M.}~\bibnamefont
  {K{\ae}rn}}\ and\ \bibinfo {author} {\bibfnamefont {M.}~\bibnamefont
  {Menzinger}},\ }\href {https://doi.org/10.1103/PhysRevE.61.3334} {\bibfield
  {journal} {\bibinfo  {journal} {Physical Review E - Statistical Physics,
  Plasmas, Fluids, and Related Interdisciplinary Topics}\ }\textbf {\bibinfo
  {volume} {61}},\ \bibinfo {pages} {3334} (\bibinfo {year}
  {2000})}\BibitemShut {NoStop}%
\bibitem [{\citenamefont {McGraw}\ and\ \citenamefont
  {Menzinger}(2005)}]{McGraw2005}%
  \BibitemOpen
  \bibfield  {author} {\bibinfo {author} {\bibfnamefont {P.~N.}\ \bibnamefont
  {McGraw}}\ and\ \bibinfo {author} {\bibfnamefont {M.}~\bibnamefont
  {Menzinger}},\ }\href {https://doi.org/10.1103/PhysRevE.72.026210} {\bibfield
   {journal} {\bibinfo  {journal} {Physical Review E}\ }\textbf {\bibinfo
  {volume} {72}},\ \bibinfo {pages} {026210} (\bibinfo {year}
  {2005})}\BibitemShut {NoStop}%
\bibitem [{\citenamefont {Krause}\ \emph {et~al.}(2018)\citenamefont {Krause},
  \citenamefont {Klika}, \citenamefont {Woolley},\ and\ \citenamefont
  {Gaffney}}]{Krause2018}%
  \BibitemOpen
  \bibfield  {author} {\bibinfo {author} {\bibfnamefont {A.~L.}\ \bibnamefont
  {Krause}}, \bibinfo {author} {\bibfnamefont {V.}~\bibnamefont {Klika}},
  \bibinfo {author} {\bibfnamefont {T.~E.}\ \bibnamefont {Woolley}},\ and\
  \bibinfo {author} {\bibfnamefont {E.~A.}\ \bibnamefont {Gaffney}},\ }\href
  {https://doi.org/10.1103/PhysRevE.97.052206} {\bibfield  {journal} {\bibinfo
  {journal} {Physical Review E}\ }\textbf {\bibinfo {volume} {97}},\ \bibinfo
  {pages} {052206} (\bibinfo {year} {2018})}\BibitemShut {NoStop}%
\bibitem [{\citenamefont {Kolokolnikov}\ \emph {et~al.}(2018)\citenamefont
  {Kolokolnikov}, \citenamefont {Ward}, \citenamefont {Tzou},\ and\
  \citenamefont {Wei}}]{Kolokolnikov2018}%
  \BibitemOpen
  \bibfield  {author} {\bibinfo {author} {\bibfnamefont {T.}~\bibnamefont
  {Kolokolnikov}}, \bibinfo {author} {\bibfnamefont {M.}~\bibnamefont {Ward}},
  \bibinfo {author} {\bibfnamefont {J.}~\bibnamefont {Tzou}},\ and\ \bibinfo
  {author} {\bibfnamefont {J.}~\bibnamefont {Wei}},\ }\href
  {https://doi.org/10.1098/rsta.2018.0110} {\bibfield  {journal} {\bibinfo
  {journal} {Philosophical Transactions of the Royal Society A: Mathematical,
  Physical and Engineering Sciences}\ }\textbf {\bibinfo {volume} {376}},\
  \bibinfo {pages} {20180110} (\bibinfo {year} {2018})}\BibitemShut {NoStop}%
\bibitem [{\citenamefont {Mahoney}\ and\ \citenamefont
  {Mitchell}(2015)}]{Mahoney2015}%
  \BibitemOpen
  \bibfield  {author} {\bibinfo {author} {\bibfnamefont {J.~R.}\ \bibnamefont
  {Mahoney}}\ and\ \bibinfo {author} {\bibfnamefont {K.~A.}\ \bibnamefont
  {Mitchell}},\ }\href {https://doi.org/10.1063/1.4922026} {\bibfield
  {journal} {\bibinfo  {journal} {Chaos: An Interdisciplinary Journal of
  Nonlinear Science}\ }\textbf {\bibinfo {volume} {25}},\ \bibinfo {pages}
  {087404} (\bibinfo {year} {2015})}\BibitemShut {NoStop}%
\bibitem [{\citenamefont {Bresciani}\ \emph {et~al.}(2019)\citenamefont
  {Bresciani}, \citenamefont {Kang},\ and\ \citenamefont
  {Lee}}]{Bresciani2019}%
  \BibitemOpen
  \bibfield  {author} {\bibinfo {author} {\bibfnamefont {E.}~\bibnamefont
  {Bresciani}}, \bibinfo {author} {\bibfnamefont {P.~K.}\ \bibnamefont
  {Kang}},\ and\ \bibinfo {author} {\bibfnamefont {S.}~\bibnamefont {Lee}},\
  }\href {https://doi.org/10.1029/2018WR023508} {\bibfield  {journal} {\bibinfo
   {journal} {Water Resources Research}\ }\textbf {\bibinfo {volume} {55}},\
  \bibinfo {pages} {1624} (\bibinfo {year} {2019})}\BibitemShut {NoStop}%
\end{thebibliography}
%

\end{document}